# Effect of polymeric additives on ignition, combustion and flame characteristics and soot deposits of crude oil droplets

Gurjap Singh [1*], Mehdi Esmaeilpour [2], Albert Ratner [1]

[1] Department of Mechanical Engineering, The University of Iowa, Iowa City, IA 52242, USA
[2] College of Information Technology and Engineering, Marshall University, Huntington, WV 25755, USA

**Contact: gurjap-singh@uiowa.edu*

## Abstract:

Many oil fires have resulted from the crude oil train derailments in recent years. Given the importance of crude oil shipping by rail to the energy security of the US, it is important to consider various methods that will decrease the likelihood of crude oil catching fire in case of a crude oil derailment. Present study examines the effect of polybutadiene polymer on the combustion properties and soot deposits of Bakken and Pennsylvania crudes. Treating these crudes as multicomponent liquid fuels and polybutadiene as an additive, droplet combustion experiments were conducted with sub-millimeter sized spherical droplets suspended on very fine support fibers. Polybutadiene polymer additive of two different chain lengths has been investigated. Results show that both polymer chain length and origin of crude oil have a significant effect on various combustion properties like combustion rate, ignition delay, total combustion time, and flame stand-off ratio. Polymeric additives also change the soot deposit structure and particle size compared to the base fuel. Present research is envisioned to aid in theoretical combustion modeling of complex multicomponent liquid fuels, as well as generate interest in investigating more polymeric additives for liquid fuels.

*Number of words: 189*

## Nomenclature:

$d_0$ = initial droplet diameter
$d(t)$ = the droplet diameter at time $t$
$t$ or $\tau$ = combustion time
$k$ or $K$ = burning rate
$M_n$ = polymer chain length

## Abbreviations:

CCD: charge-coupled device
CMOS: complementary metal-oxide semiconductor
FSR: flame stand-off ratio
GC-MS: gas chromatography-mass spectrometry
PBD: Polybutadiene
PBD5k: Polybutadiene $M_n = 5{,}000$
PBD200k: Polybutadiene $M_n = 200{,}000$

## 1. Introduction:

The oil boom in the North Dakota Bakken formation has presented a logistical problem of having to ship the said crude to oil refineries. The proposed Keystone XL pipeline, meant to connect the Bakken oilfields to the already-existing Keystone oil distribution system [1] has been delayed indefinitely. The shortfall in pipeline transportation has been taken up by rail. From 2010 to 2015, crude oil shipments from the Midwest to the rest of the US have steadily increased [2]. Rail infrastructure has become critical to the energy security of the United States, with rail crude transport supplying more than half the feedstock of East Coast refineries [3].

However, this extra shipping load has placed an unprecedented stress on the aging US rail infrastructure. In recent years, many oil train derailments and crashes have occurred, which generally result in devastating oil fires and a loss of life and property [4][5][6]. Such oil fires are of special relevance to the very sweet and light Bakken crude oil, which contains a large amount of easy to vaporize, easy to burn light ends [7].

It is an expensive and time-consuming proposition to upgrade US rail infrastructure to promote safer crude oil transport. This manuscript explores polymeric additives as a stopgap measure to control the burning behavior of crude in case of a crude train derailment and spillage. Previous research has investigated the addition of long chained polymers to diesel and its blends to suppress mist formation and splashing [8]. It has been found that adding long chain polymers to diesel and Jet-A droplets [9] and their surrogate blends [10], [11] slows down their burning rate and increases their ignition delay. Most notably, polymers are already being used in crude oil pipelines as drag reducing agents [12], [13].

Very few systematic crude oil combustion studies are available currently, and those that are available generally explore pool fire properties of crude oils in relation to in-situ burning. Previous work by the authors has explored the droplet combustion properties of crude oils from various US oil production regions (Bakken, Colorado, Pennsylvania, Texas) [14], as well as combustion property modification of Bakken crude oil using nano-additives [15]. It was found that Bakken, Colorado, and Pennsylvania crude oil droplet combustion rates were comparable, but Texas crude oil burned with high microexplosion intensity. Present work similarly treats Pennsylvania and Bakken crudes as multicomponent liquid fuels and aims to establish the modification to their droplet combustion properties and soot residue properties when polybutadiene (PBD) polymer of two chain lengths (5,000 and 200,000) is blended to them in various proportions.

A well-established, bench-top method has been used for present work, which has previously used to establish various combustion properties for isolated spherical droplets of liquid fuels such as combustion rate [9], [10], [14]–[18], ignition delay [14], [15], [17], and total combustion time [14], [15], [17]. It uses only a small amount of sample, promotes spherical symmetry, and yields conveniently usable data for further analysis.

The crude oil used in present work was obtained from the Northeast oil production region (Pennsylvania) and the Rocky Mountain oil production region (Bakken) of the United States. Although significant variety exists in crude oil properties, **Table 1** presents typical properties of Bakken crude [7] and Pennsylvania crude [19]. Additionally, Appendix A presents gas chromatography-mass spectrometry (GC-MS) data for both crude oils.

**Table 1.** Typical properties of crudes tested

| Property/Crude Oil | Bakken crude | Pennsylvania crude |
|---|---|---|
| **Initial Boiling Point [$^0$C]** | 21 | 30 |
| **Specific Gravity** | 0.815 | 0.832 |

Polybutadiene (PBD) was obtained from Sigma Aldrich. Formed from the polymerization of 1,3-butadiene ($H2C = C = CH - CH2$) monomer, PBD is a commercially available, long chain polymer with a several applications, such as making tires, golf balls, and toys. One PBD variety that was tested had chain length $M_n = \sim 5{,}000$ and is referred to as PBD5k in this manuscript. The other variety had chain length $M_n = \sim 200{,}000$ and is referred to as PBD200k in this work. Both PBD5k and PBD200k are readily combustible, but PBD5k is a very viscous liquid whereas PBD200k is a rubbery solid.

## 2. Experimental Method:

The experimental apparatus has previously been previously detailed in the work presented by Singh et al. [14], [17], [15] which in turn was inspired from Avedisian and Callahan [20] and Bae and Avedisian [21]. **Figure 1** shows the schematic of the experiment. A short summary is as follows: a sub-millimeter sized droplet of the fuel is generated using a microsyringe and suspended using three 16 µm silicon-carbide (SiC) fibers. These are fixed between six posts as shown and meet in the center. Solenoids control semi-circular hot wires of 1 mm diameter, which are positioned close to the droplet. At an appropriate time (generally 500ms) the solenoids are activated and withdraw the hot wire away. These hot wires are made of 36-gauge Kenthal with a resistance of 4Ω each and are connected to a power supply of 15 V to make them glow red-hot. Combined, they produce energy equivalent to 14.06 J in a typical experiment where they are energized for 500ms.

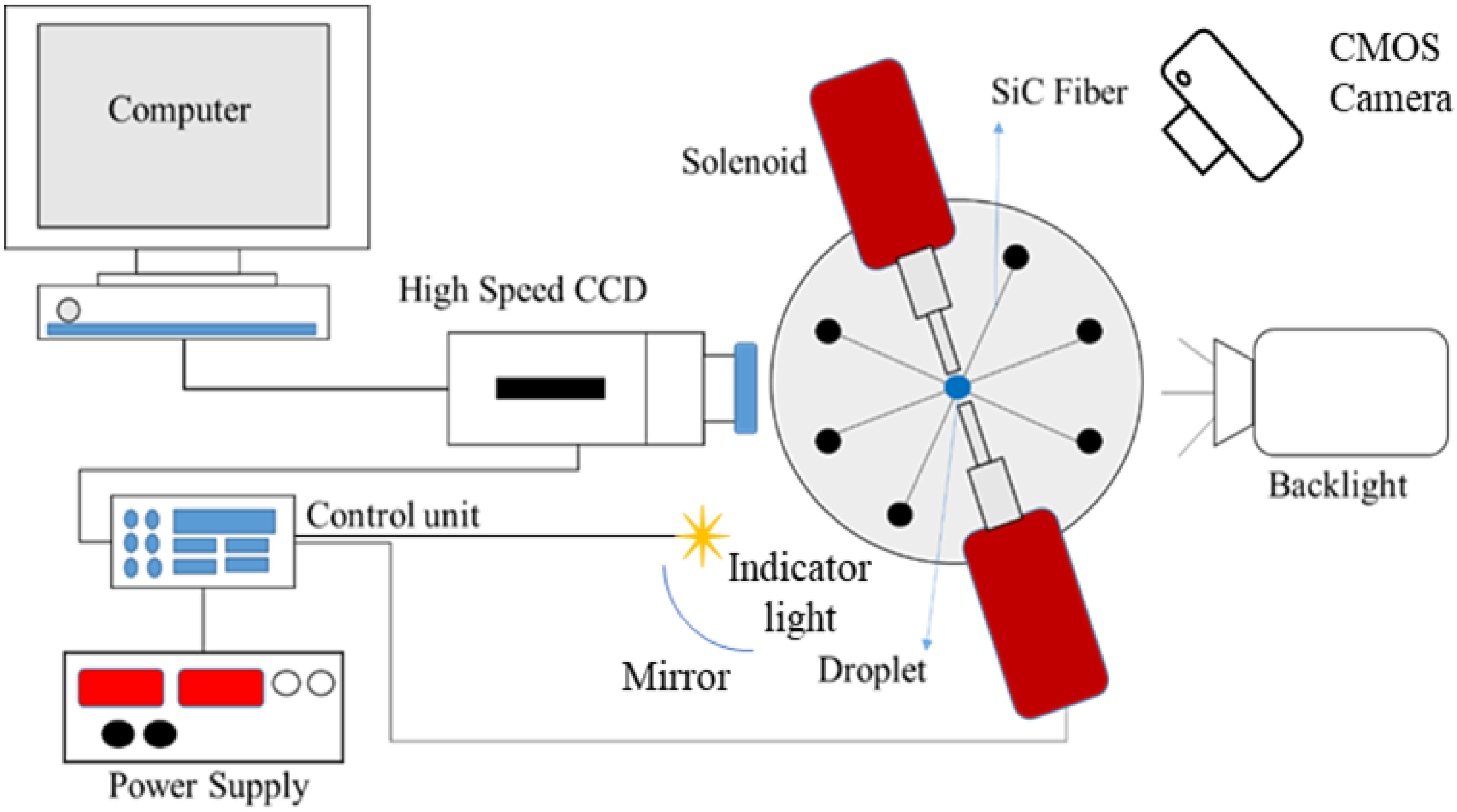

**Figure 1.** Experimental setup schematic showing the location of the fuel droplet and various components [14]

A microcontroller (Arduino Uno) controls the time for which the hot wires are energized and when they retract. In turn, the microcontroller is controlled by a signal from the high-speed camera. The system acts in sync: when the camera starts to acquire data, the microcontroller is activated, and it controls the hot wires and the solenoid in turn.

The high-speed camera is a black and white charge-coupled device (CCD) IDT X-StreamVision XS-3 (IDT Vision, Pasadena, CA, USA) camera operated at 1000 frames per second and fitted with a 105 mm lens (Nikon AF Micro-Nikkor-F/2.8, Tokyo, Japan). Back-lighting is provided by a single bright white LED at 3.3 V. The ignition and combustion process is simultaneously recorded through a magnifying concave mirror (4.0" diameter, 9.0" focal length) by a complementary metal-oxide semiconductor (CMOS) Casio EXILIM Pro EX-F1 (Casio, Tokyo, Japan) high-speed camera operating at 600 frames/second.

The CCD camera generates 8-bit grayscale image sequence with 948 x 592 pixels. The open-source digital image processing software ImageJ/Fiji [22]–[24] was used to post-process the images captured by the CCD camera and is used to calculate the combustion rate. The CMOS camera generates 8-bit RGB video sequence with 432 x 192 pixels, which us post-processed using MATLAB ® (MathWorks, Natick, MA, USA) to generate time-series RGB images, which are then used to calculate ignition delay and total combustion time.

Post-processing of CCD camera images yields the evolution of diameter $d$ (and therefore area $\pi d^2/4$) with time $\tau$ for a given droplet. After normalizing area and $\tau$ with squared diameter of initial droplet $d_0^2$, the evolution of normalized area $(d/d_0)^2$ vs modified time $\tau/d_0^2$ was determined. Moving average was used to reduce data, which was then further post-processed to determine combustion or burning rates.

To set a base reference, pure crude was burned with no PBD added. Polymer was blended into the crude oil in 0.5%, 1%, 2%, 3% and 4% (w/w) concentrations using a standard lab stir plate for 24 hours. No heat was used, and the stir speed was kept consistent across all experiments. Once prepared, the fuel blend was immediately used, and was returned to the stir plate between experiments. This was done to keep the fuel blend homogeneous and to prevent any possible polymer strand agglomeration.

At least 6 experiments were performed for each fuel blend, 5 with full LED backlight and 1 in low light conditions. CCD images from the low-light experiment were used to determine the flame stand-off ratio. CCD images from back-lit experiments were used for plotting time-dependent normalized area to determine combustion rate as outlined above. After post processing, the average of the combustion rates from these experiments yielded the combustion rate of that particular blend, with the standard deviation serving as the error. CMOS camera images for all 6 experiments for a given fuel blend, extracted as outlined in above paragraphs, were used to determine average ignition delay and total combustion time for each fuel blend with standard deviation serving as the error.

## 3. Results and Discussion:

**Figure 2 a-f** shows the CCD camera image sequence of a Bakken crude-2% PBD5k blend droplet as it goes through various combustion regimes. **Figure 3 a-e**, **Figure 4 a-e** shows single droplet combustion regimes of various crudes and their blends, representing normalized area evolution $(d(t)/d_0)^2$ with modified time $(\tau/d_0)^2$. As reported before in literature [14] [15] the droplet combustion regime of the crude oils can divided into four categories. As the droplet is heated up, it thermally expands and the normalized droplet area is seen to increase, which is termed as Zone I or ignition delay. The end of ignition delay is marked appearance of a flame and the onset of combustion. For some time, a steady combustion regime is observed in Zone II. The droplet heats up even furhter in this zone, which causes the low-boiling crude fractions to preferentially boil out and cause violent microexplosions in Zone III. This mechanism has been detailed previously in the work of Singh *et al.* [14]. As these are exhausted, the microexplosion intensity decreases and the droplet burns in a steadier combustion regime in Zone IV, for which a meaningful combustion rate can be calculated. For all PBD-crude blends, an extra zone (Zone V) follows at the end of Zone IV where pure polymer or its residue is seen to burn in a very steady combustion regime, and at a markedly different combustion rate compared to Zone IV. Combustion rates for both Zone IV and Zone V have been calculated in this work.

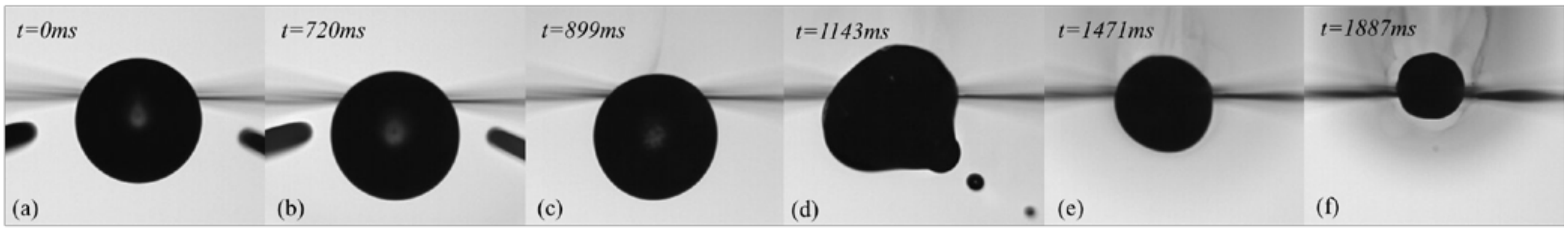

**Figure 2.** Time evolution of a Bakken crude-PBD5k 2% w/w blend droplet. (a) Initial configuration showing spherical fuel droplet deployed on suspending wires and surrounded by heating coils, (b) Droplet in Zone I: beginning of combustion after ignition delay, (c) Droplet in Zone II: steady combustion regime with heating coils retracted, (d) Droplet in Zone III: violent microexplosion combustion regime, (e) Droplet in Zone IV: steady combustion with low intensity microexplosion combustion regime, (f) Droplet in Zone V: polymer combustion regime

Because of the highly multicomponent nature of the crudes (see **Appendix A**) many microexplosions throughout the combustion regime can be seen in crude oil and PBD-crude blends. A microexplosion is characterized by a rapid spike in the normalized droplet area plot, which is followed by a rapid dip. The spike represents vapor build-up inside the droplet, causing the droplet to puff. The dip occurs when the vapor is released from the droplet surface into the flame and causing a violent fragmentation of the droplet, which results in the aforementioned rapid dip in the normalized dropelt area. Adding both PBD5k and PBD200k at low concentrations reduces microexplosions but at higher concentrations the microexplosion intensity increases (**Figure 3 a-e**, **Figure 4 a-e**). However, at higher polymer concentrations PBD200k causes more violent microexplosions compared to PBD5k.

The following sections discuss combustion properties for crudes and their polymer blends such as combustion or burning rate, ignition delay, total combustion time, and flame stand-off ratio. Comparison of the soot deposits of these fuels, taken using a scanning electron microscope, is also presented.

(a)

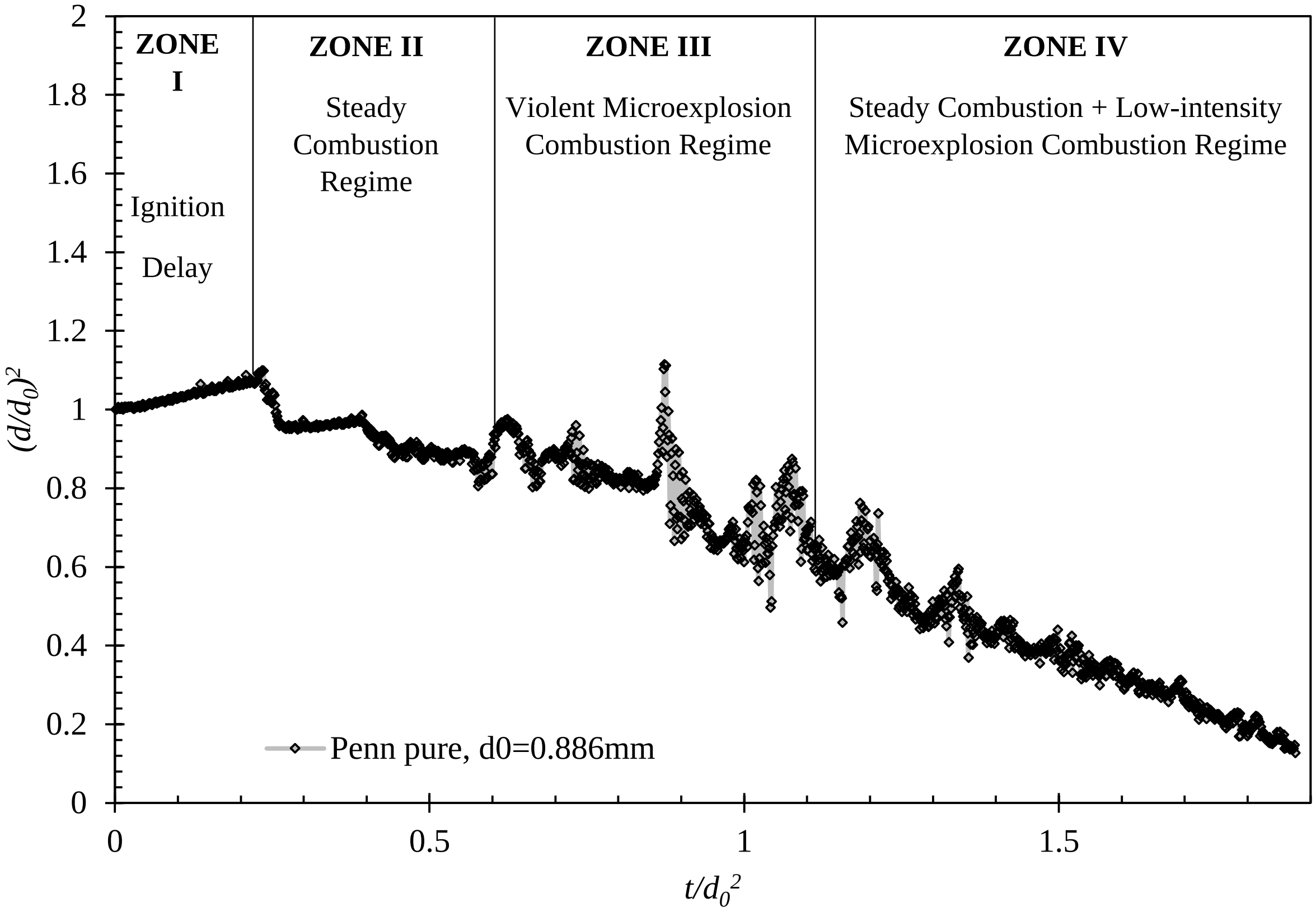

(b)

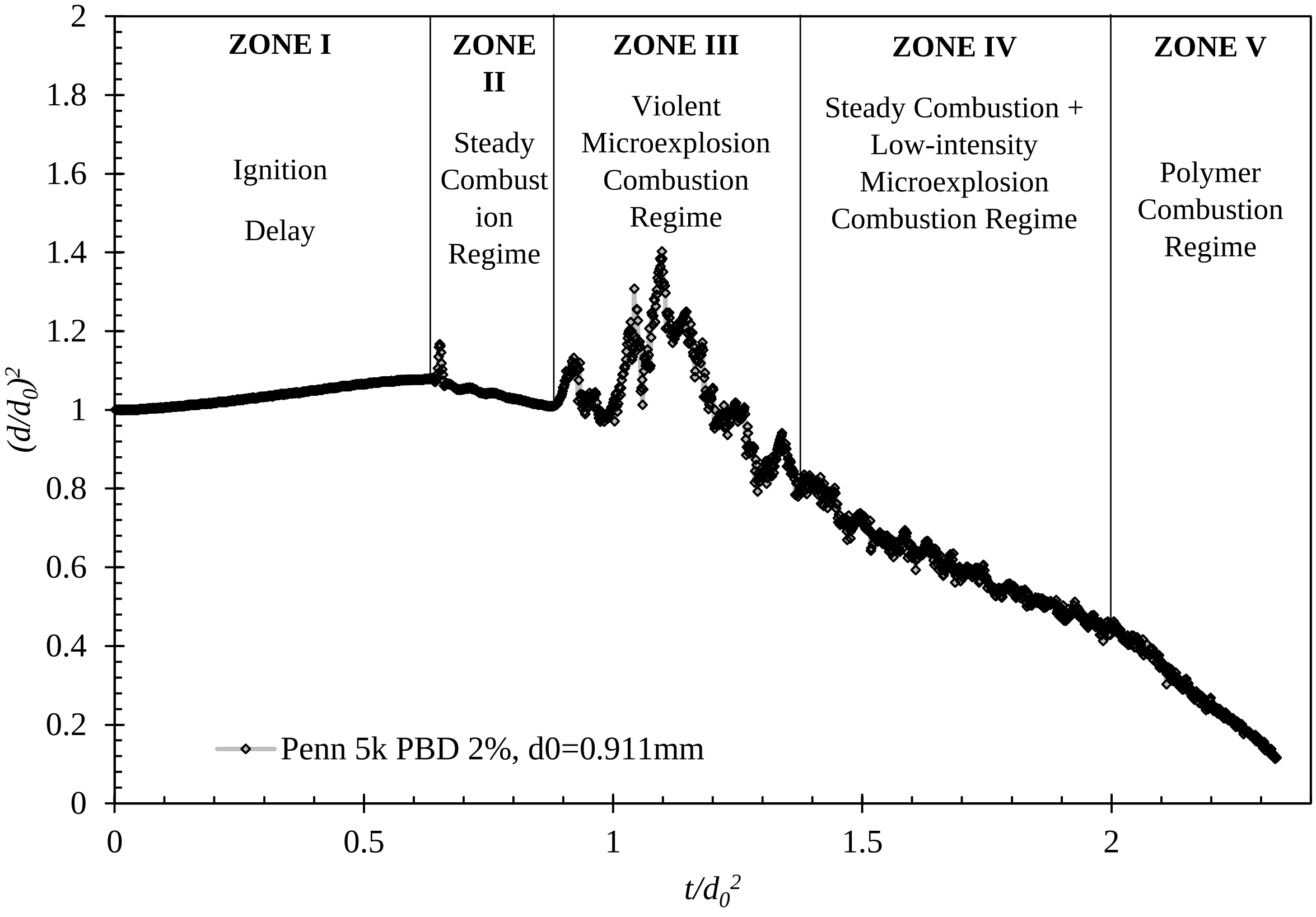
ZONE I
Ignition
Delay
ZONE II
Steady Combustion Regime
ZONE III
Violent Microexplosion Combustion Regime
ZONE IV
Steady Combustion + Low-intensity Microexplosion Combustion Regime
ZONE V
Polymer Combustion Regime
Penn 5k PBD 2%, d0=0.911mm
$(d/d_0)^2$
$t/d_0^2$

(c)

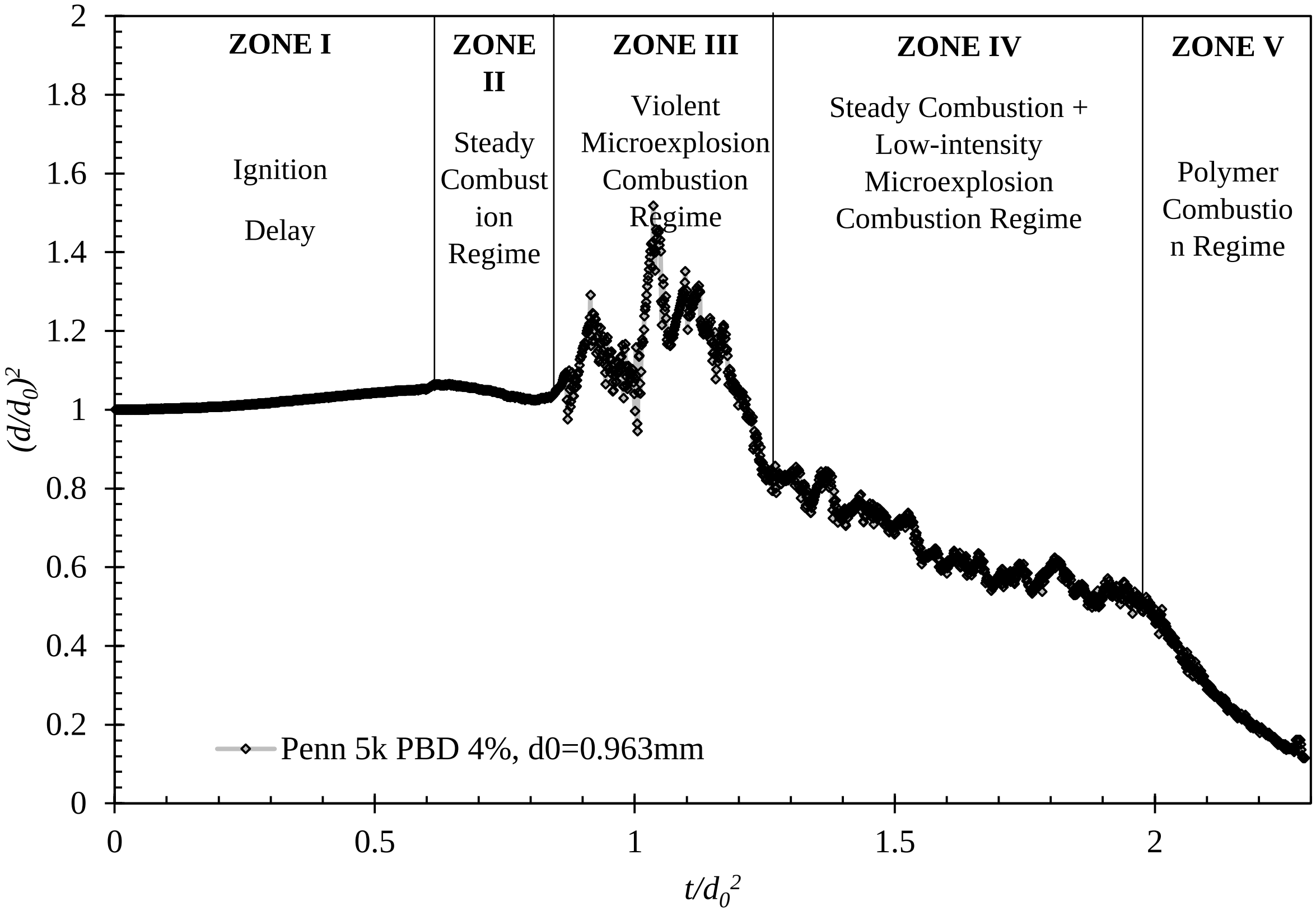
ZONE I
Ignition
Delay
ZONE II
Steady Combustion Regime
ZONE III
Violent Microexplosion Combustion Regime
ZONE IV
Steady Combustion + Low-intensity Microexplosion Combustion Regime
ZONE V
Polymer Combustion Regime
Penn 5k PBD 4%, d0=0.963mm
$(d/d_0)^2$
$t/d_0^2$

(d)

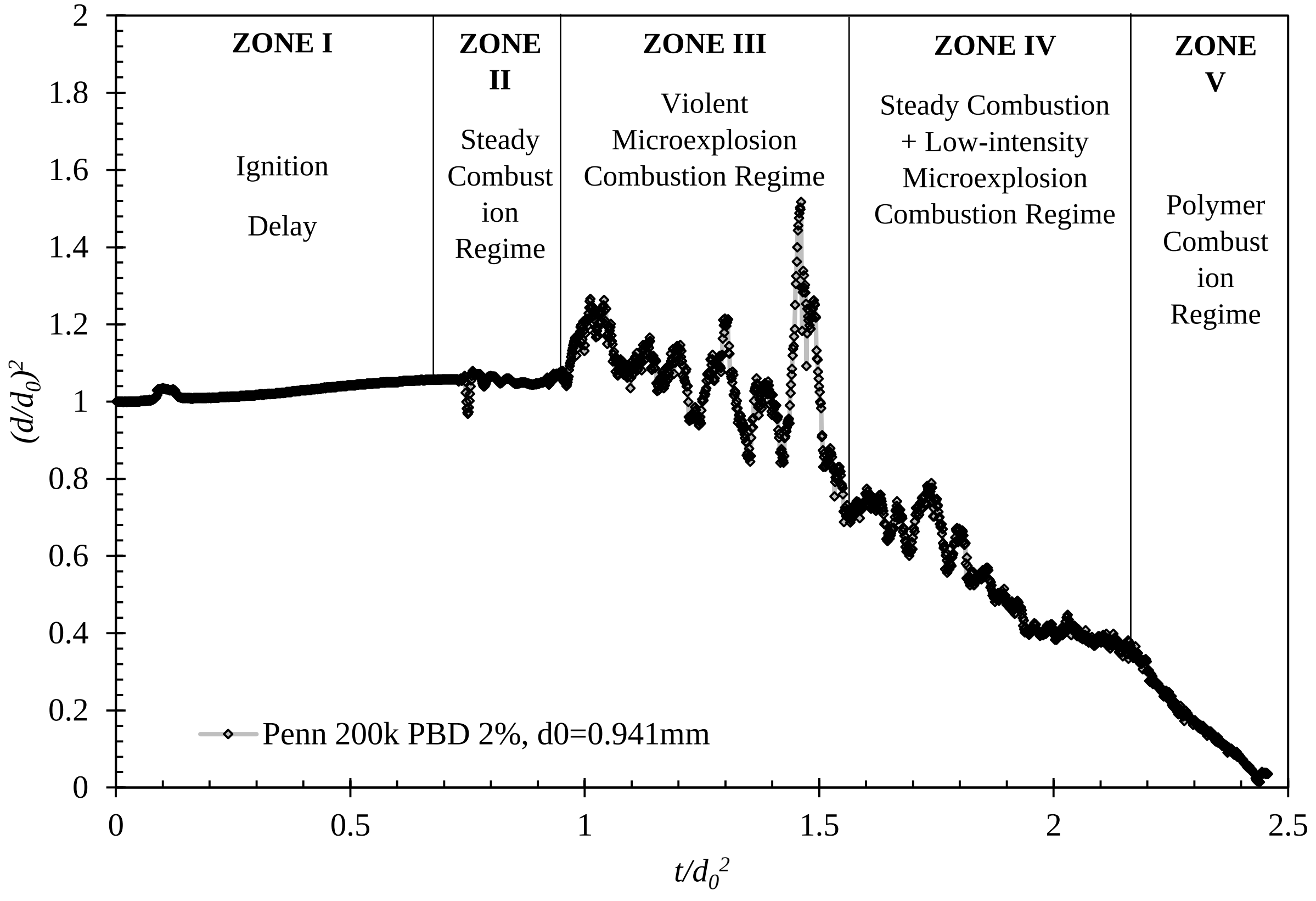
ZONE I
Ignition
Delay
ZONE II
Steady Combustion Regime
ZONE III
Violent Microexplosion Combustion Regime
ZONE IV
Steady Combustion + Low-intensity Microexplosion Combustion Regime
ZONE V
Polymer Combustion Regime
Penn 200k PBD 2%, d0=0.941mm
$(d/d_0)^2$
$t/d_0^2$
0
0.2
0.4
0.6
0.8
1
1.2
1.4
1.6
1.8
2
0
0.5
1
1.5
2
2.5

(e)

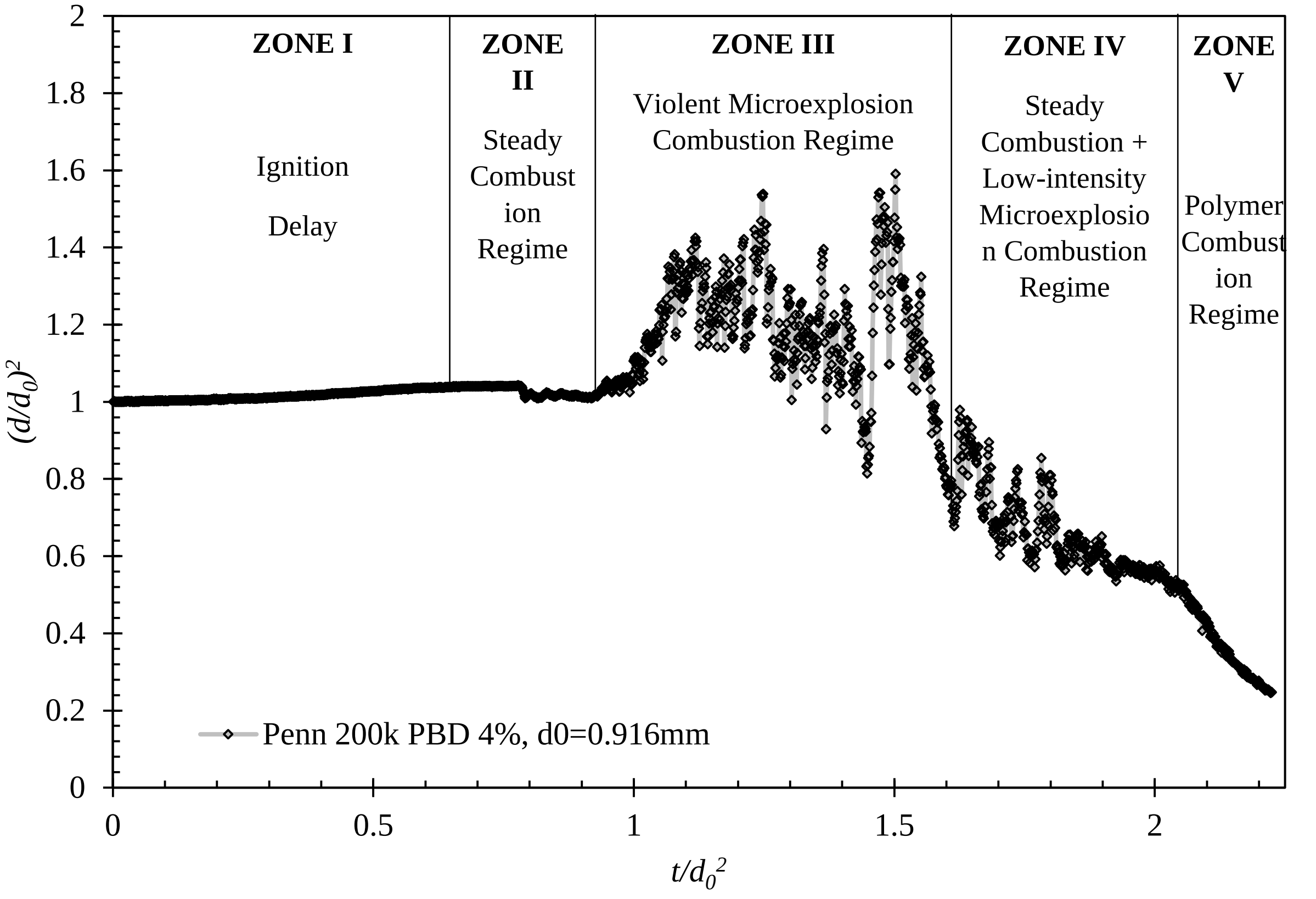
ZONE I
Ignition
Delay
ZONE II
Steady Combustion Regime
ZONE III
Violent Microexplosion Combustion Regime
ZONE IV
Steady Combustion + Low-intensity Microexplosion Combustion Regime
ZONE V
Polymer Combustion Regime
Penn 200k PBD 4%, d0=0.916mm
$(d/d_0)^2$
$t/d_0^2$
0
0.2
0.4
0.6
0.8
1
1.2
1.4
1.6
1.8
2
0
0.5
1
1.5
2

**Figure 3.** Comparison of various combustion regimes as seen in typical droplet combustion experiments for (a) Pure Pennsylvania crude [14], (b) 2% PBD5k-Penn blend, (c) 4% PBD5k-Penn blend, (d) 2% PBD200k-Penn blend, (e) 4% PBD200k-Penn blend. Note the lack of Zone V in pure Penn crude burning characteristics. Polymer-crude blends at other concentrations follow a similar trend.

(a)

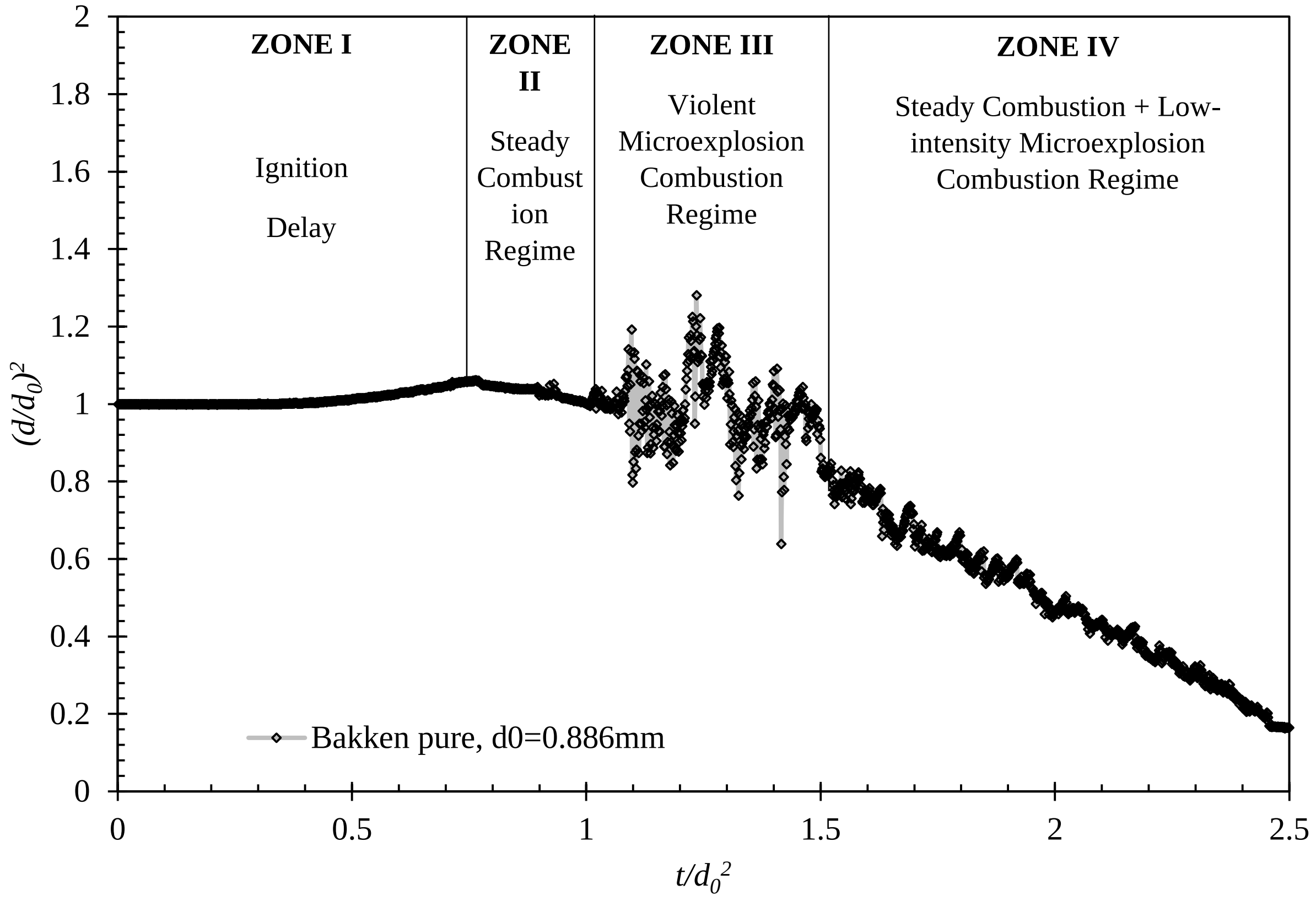

(b)

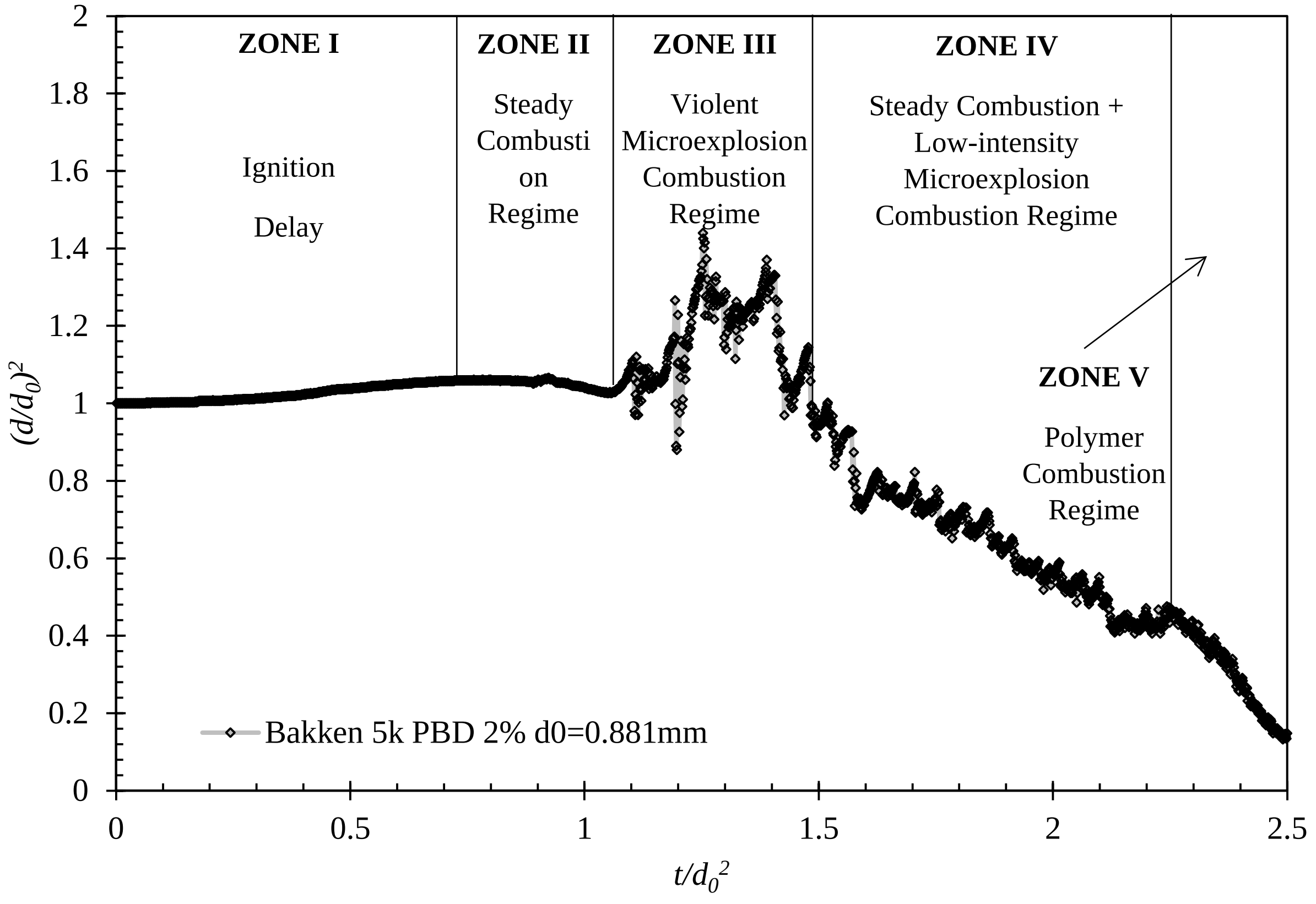

(c)

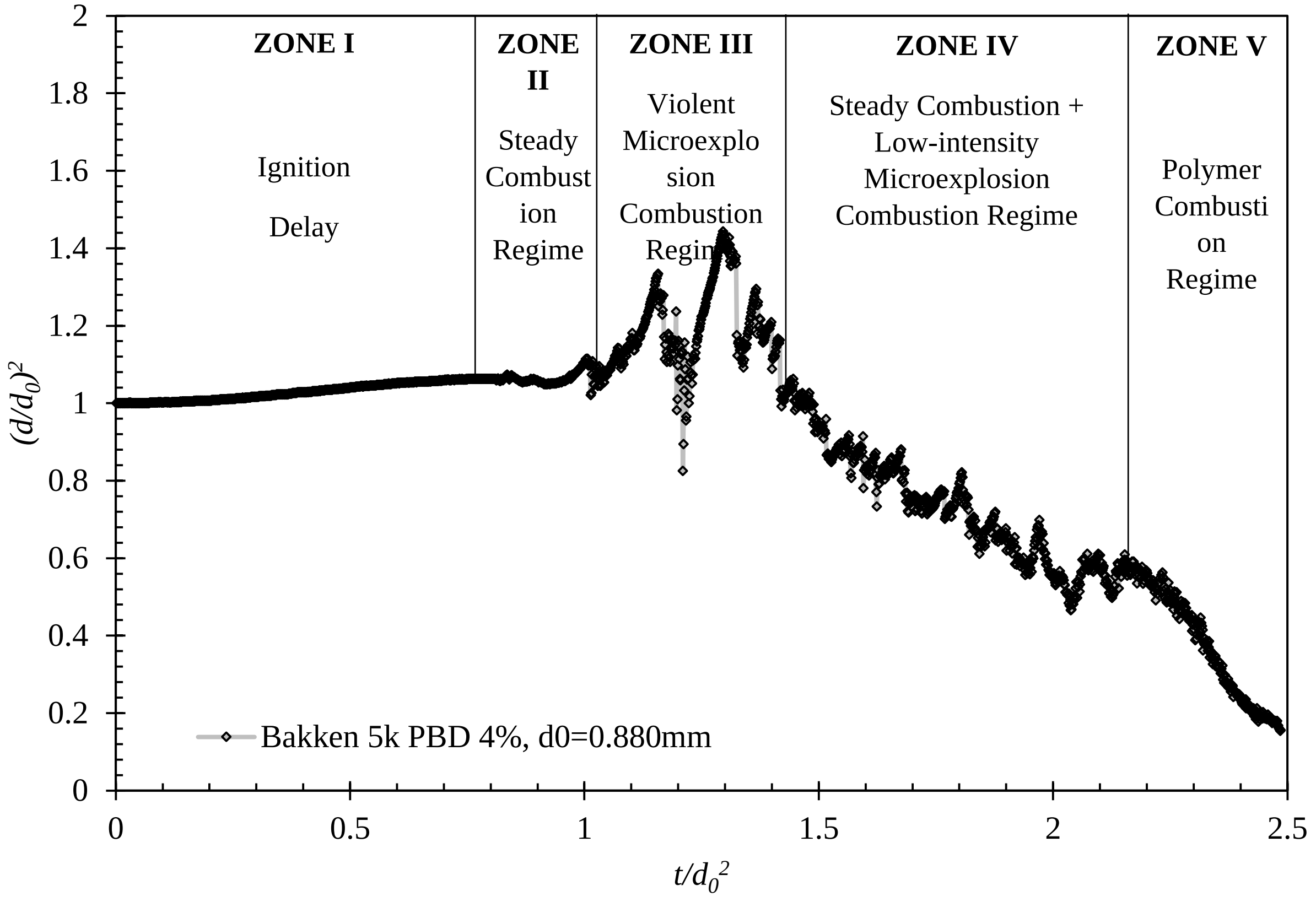

(d)

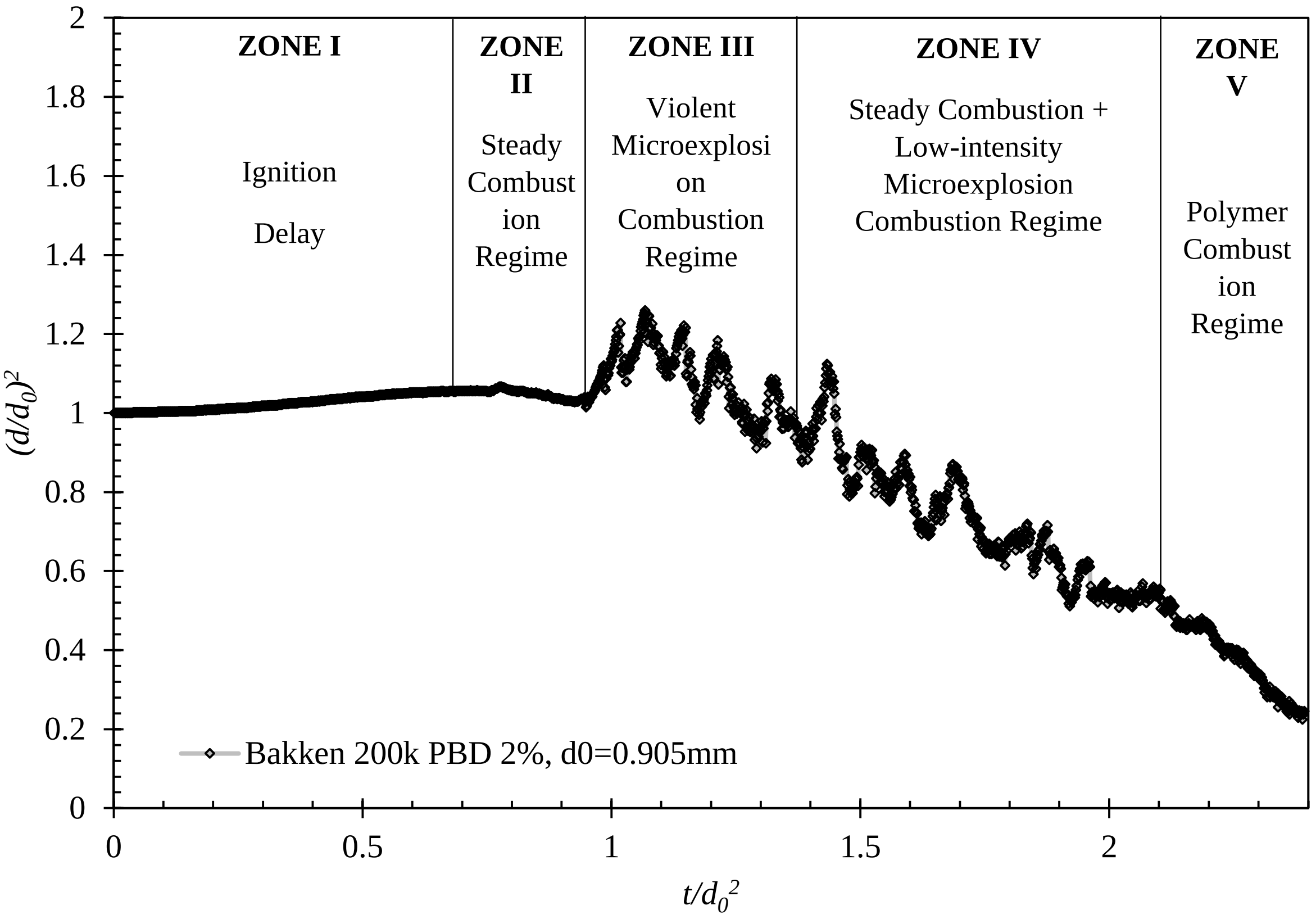

ZONE I
Ignition
Delay
ZONE II
Steady Combustion Regime
ZONE III
Violent Microexplosion Combustion Regime
ZONE IV
Steady Combustion + Low-intensity Microexplosion Combustion Regime
ZONE V
Polymer Combustion Regime
Bakken 200k PBD 2%, d0=0.905mm
$(d/d_0)^2$
$t/d_0^2$

(e)

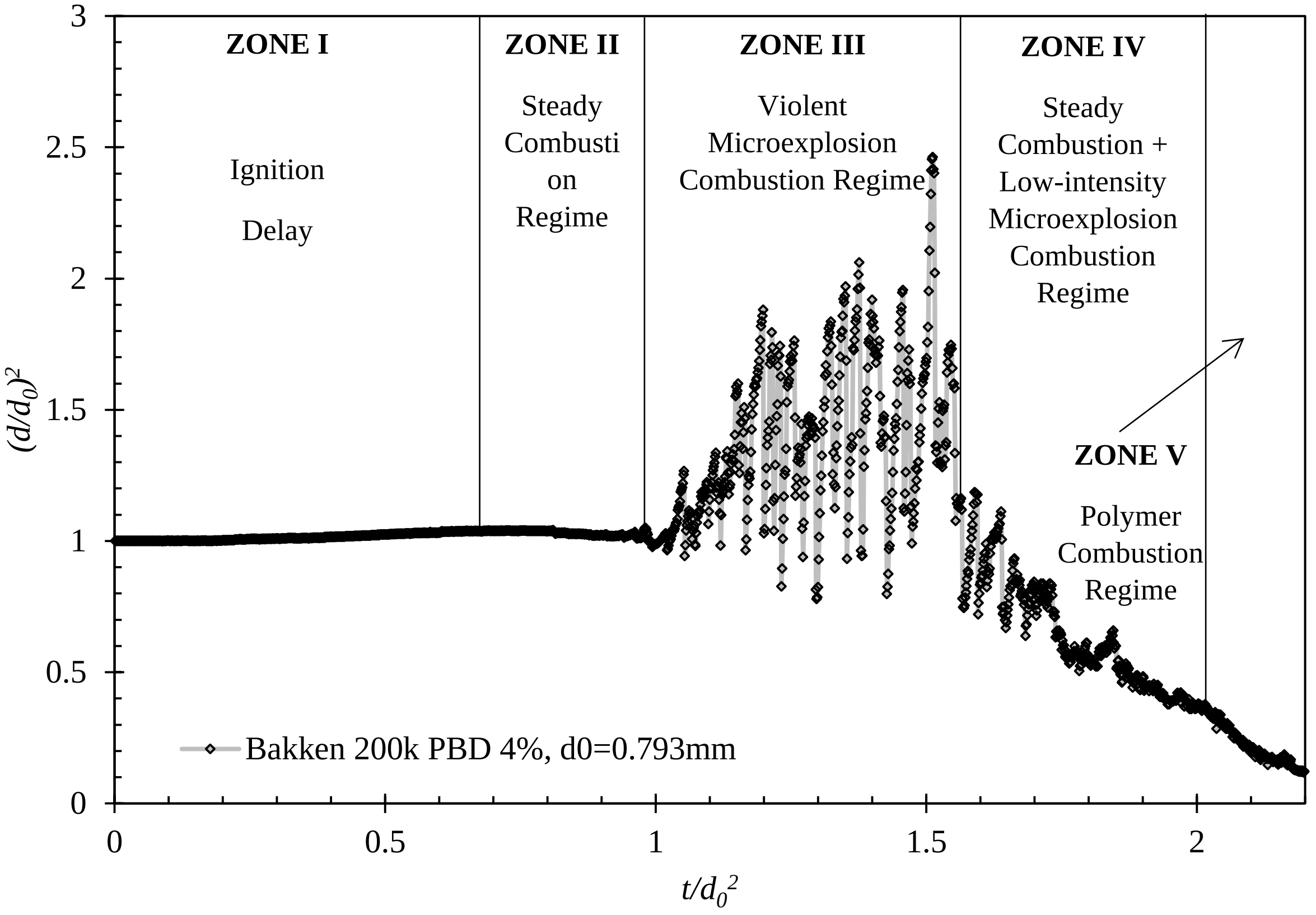

ZONE I
Ignition
Delay
ZONE II
Steady Combustion Regime
ZONE III
Violent Microexplosion Combustion Regime
ZONE IV
Steady Combustion + Low-intensity Microexplosion Combustion Regime
ZONE V
Polymer Combustion Regime
Bakken 200k PBD 4%, d0=0.793mm
$(d/d_0)^2$
$t/d_0^2$

**Figure 4.** Comparison of various combustion regimes as seen in typical droplet combustion experiments for (a) Pure Bakken crude [14], (b) 2% PBD5k-Bakken blend, (c) 4% PBD5k-Bakken blend, (d) 2% PBD200k-Bakken blend, (e) 4% PBD200k-Bakken blend. Note the lack of Zone V in pure Bakken crude burning characteristics. Polymer-crude blends at other concentrations follow a similar trend.

### 3.1 *Burning rate of fuel droplets*

Combustion characteristics in Zone IV and Zone V follow the classical $d^2$ law [25][26]:

$$\frac{d(t)^2}{d_0^2} = 1 - k(\frac{t}{d_0^2}) \tag{1}$$

where $d_0$ is the initial droplet diameter (mm), $d(t)$ is the droplet diameter (mm), $t$ denotes the combustion time (sec), and $k$ is the burning rate (mm$^2$/s).

Zone IV and Zone V combustion rates are substantially different and show very different trends as polymer concentrations increase. Zone IV combustion rate is referred to as $k_1$and Zone V combustion rate is referred to $k_2$in this manuscript. $k_1$and $k_2$for different fuels and fuel blends has been compared in this section, but since both pure Pennsylvania and pure Bakken crudes lack a Zone V, in general only a comparison for $k_1$can be made across all fuels and all fuel blends. Therefore, burning or combustion rate in this manuscript, unless otherwise noted, refers to $k_1$.

**Figure 5** shows the $k_1$burning rate trend for Pennsylvania crude oil as PBD5k and PBD200k is added, with **Figure 6** showing the average change in $k_1$. For Pennsylvania crude, PBD5k is seen to be highly effective at reducing burning rate $k_1$with large combustion rate decrease noted at all concentrations, with the largest decrease of 26% seen at 3%w/w PBD5k. PDB200k shows a significant reduction of 10% in burning rate $k_1$at 1%w/w, but a moderate to large increase in $k_1$at all other blends tested with Pennsylvania crude, with the largest increase (27%) noted at 3%w/w.

(a)

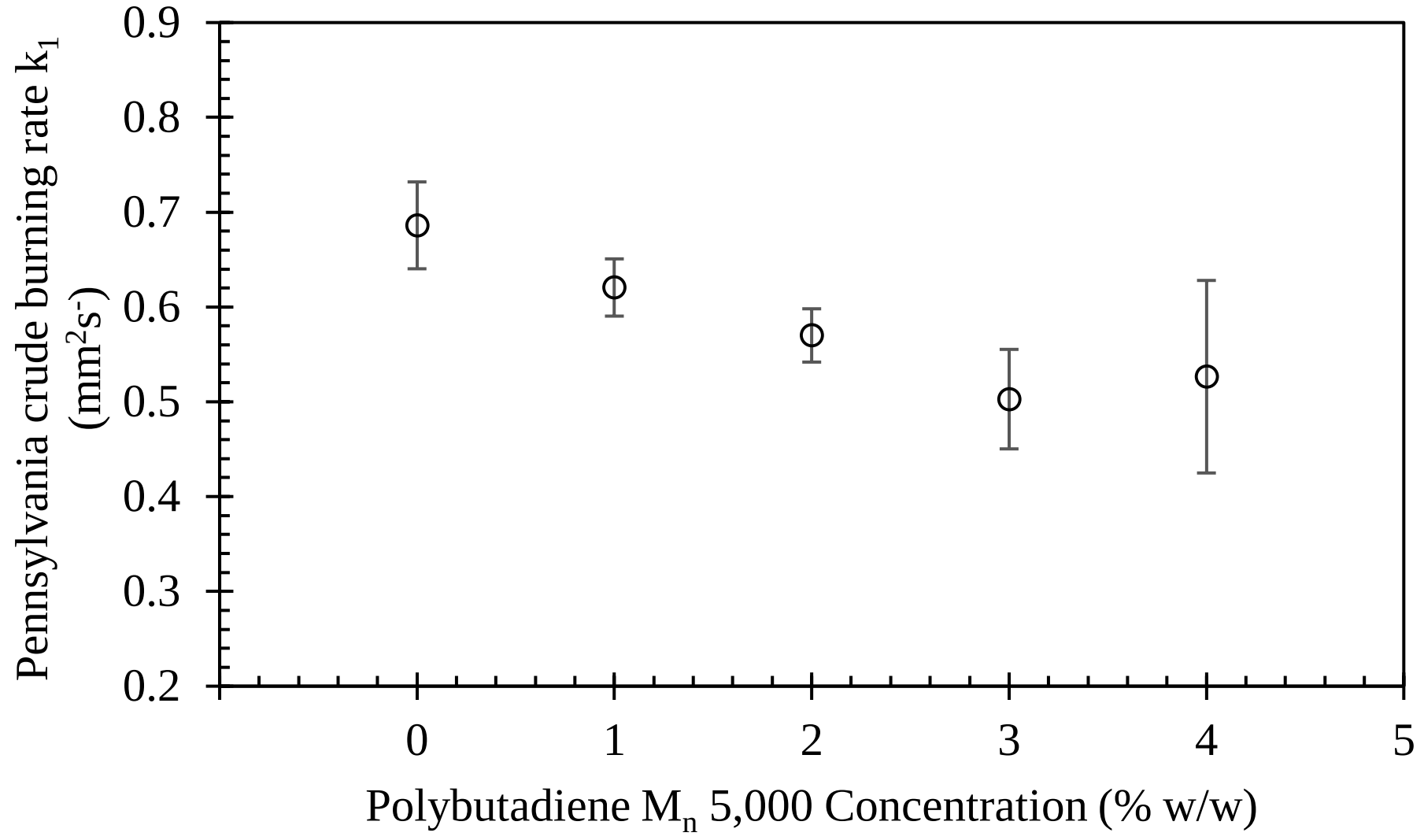

(b)

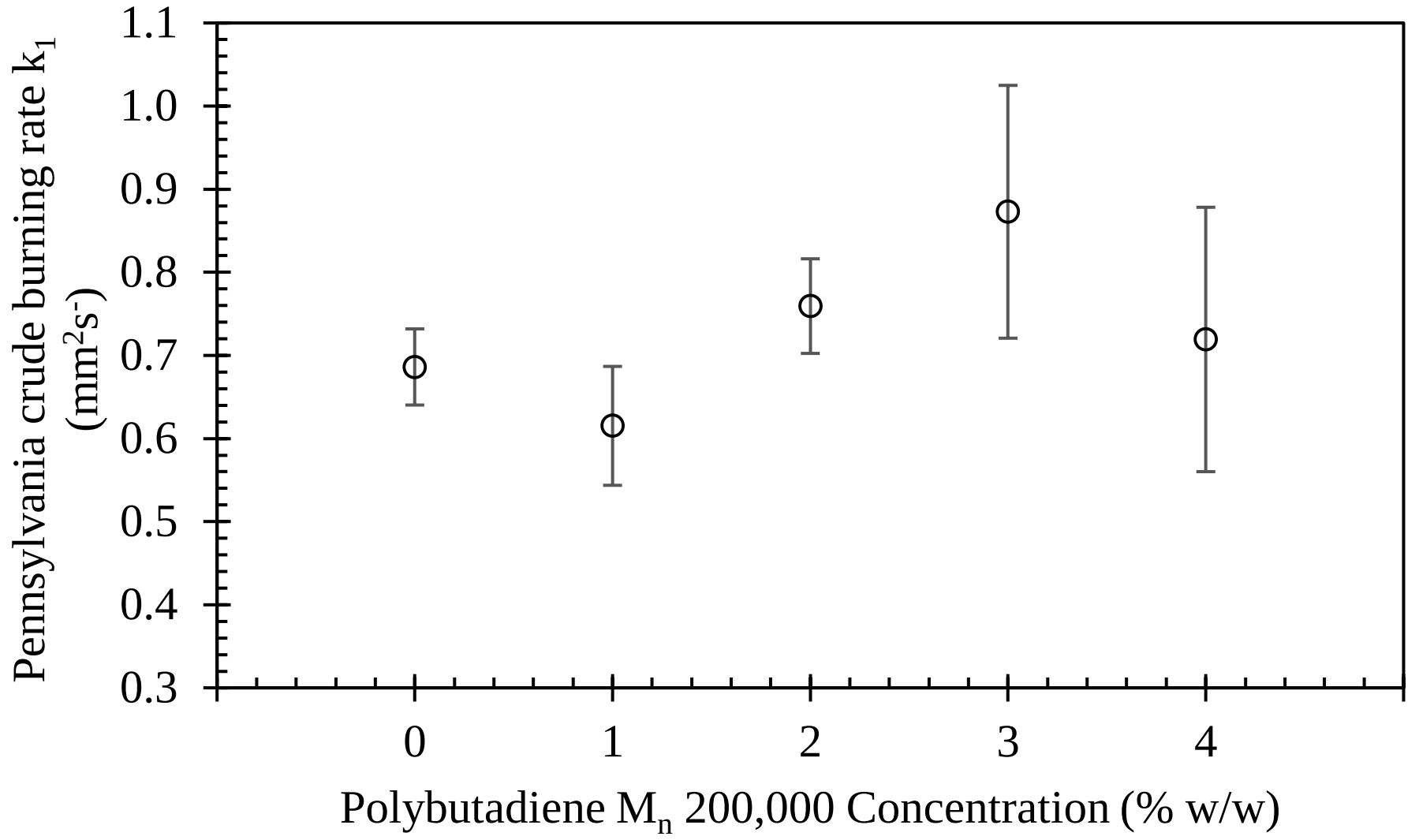

**Figure 5.** Effect of (a) 5k and (b) 200k chain length PBD on Pennsylvania crude oil combustion rate $k_1$. Error bars show the standard deviation of all experiments.

(a)

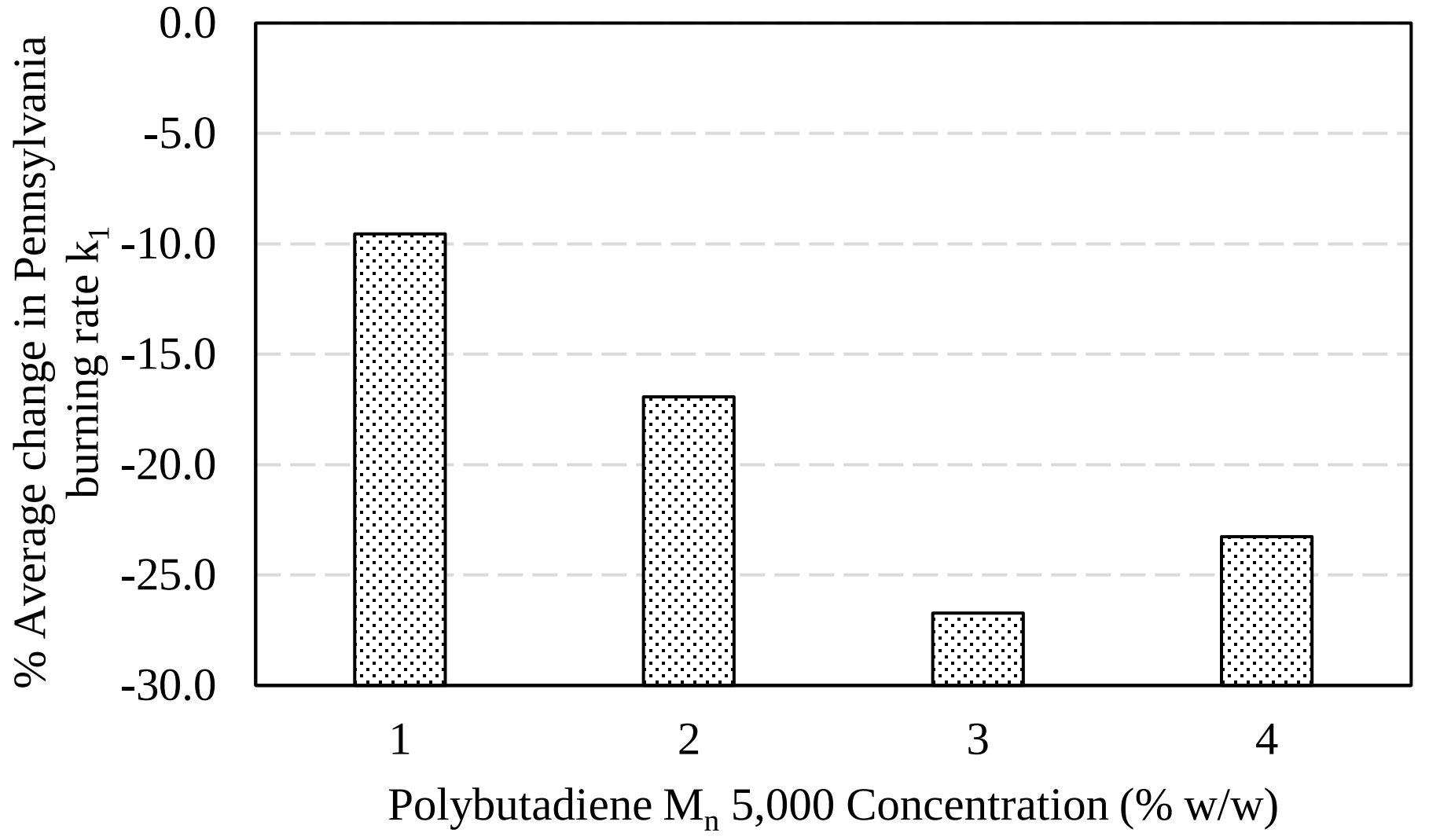

(b)

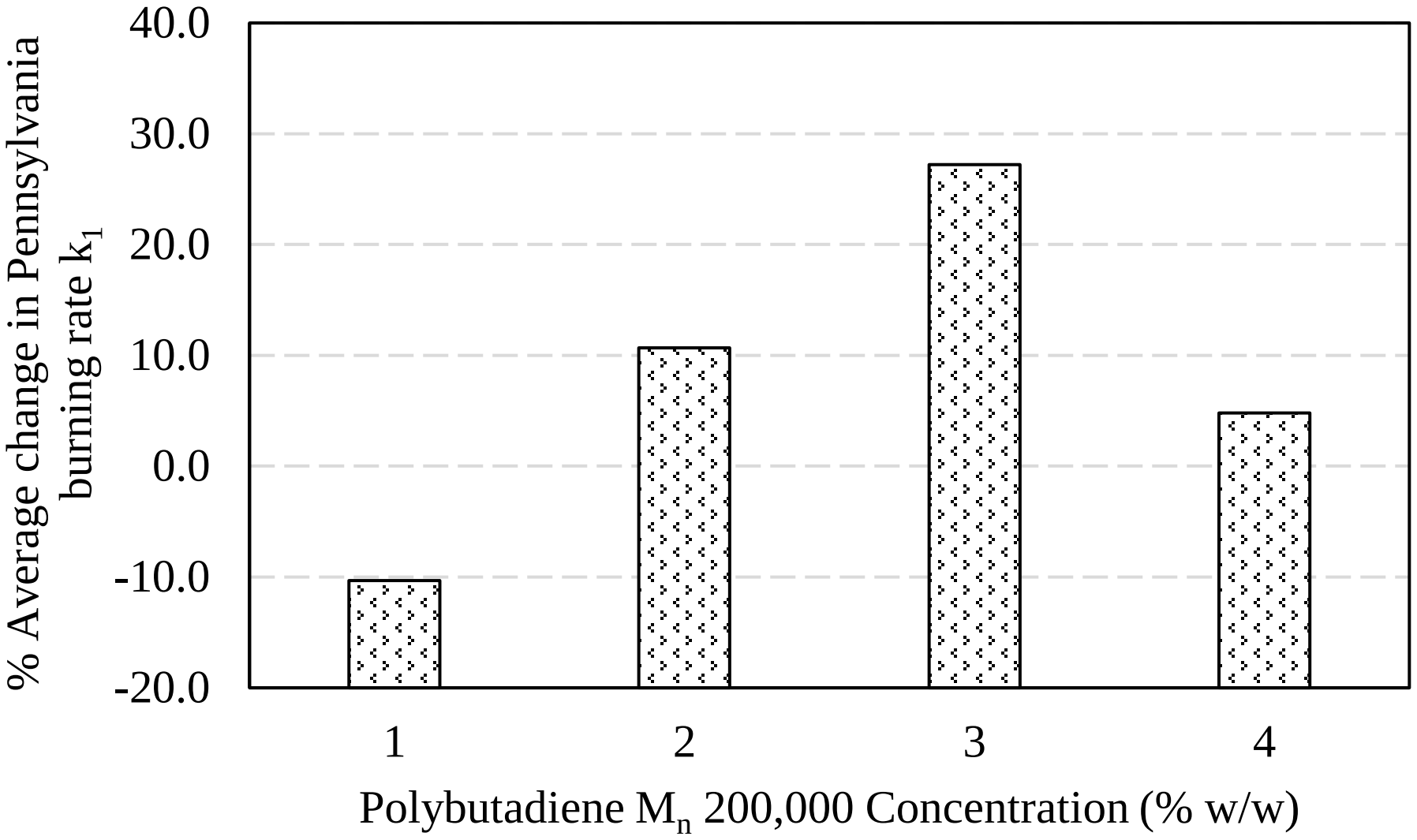

**Figure 6**. Effect of (a) 5k and (b) 200k chain length PBD on Pennsylvania crude oil average combustion rate $k_1$.

**Figure 7** shows the change in burning rate $k_1$of Bakken crude oil droplets as PBD5k and PBD200k are added at different concentrations. **Figure 8** shows the average changes in $k_1$. Like PBD5k-Pennsylvania crude blends, a decrease in burning rate $k_1$is noted for all PBD5k-Bakken concentrations with the largest decrease (12%) at 2%w/w, except at 3%w/w where an increase of 5% is noted. Again, similar to PBD200k-Pennsylvania blends, PDB200k generally causes an increase in Bakken crude burning rate $k_1$, with the largest increase of 140% noted at 3%w/w.

(a)

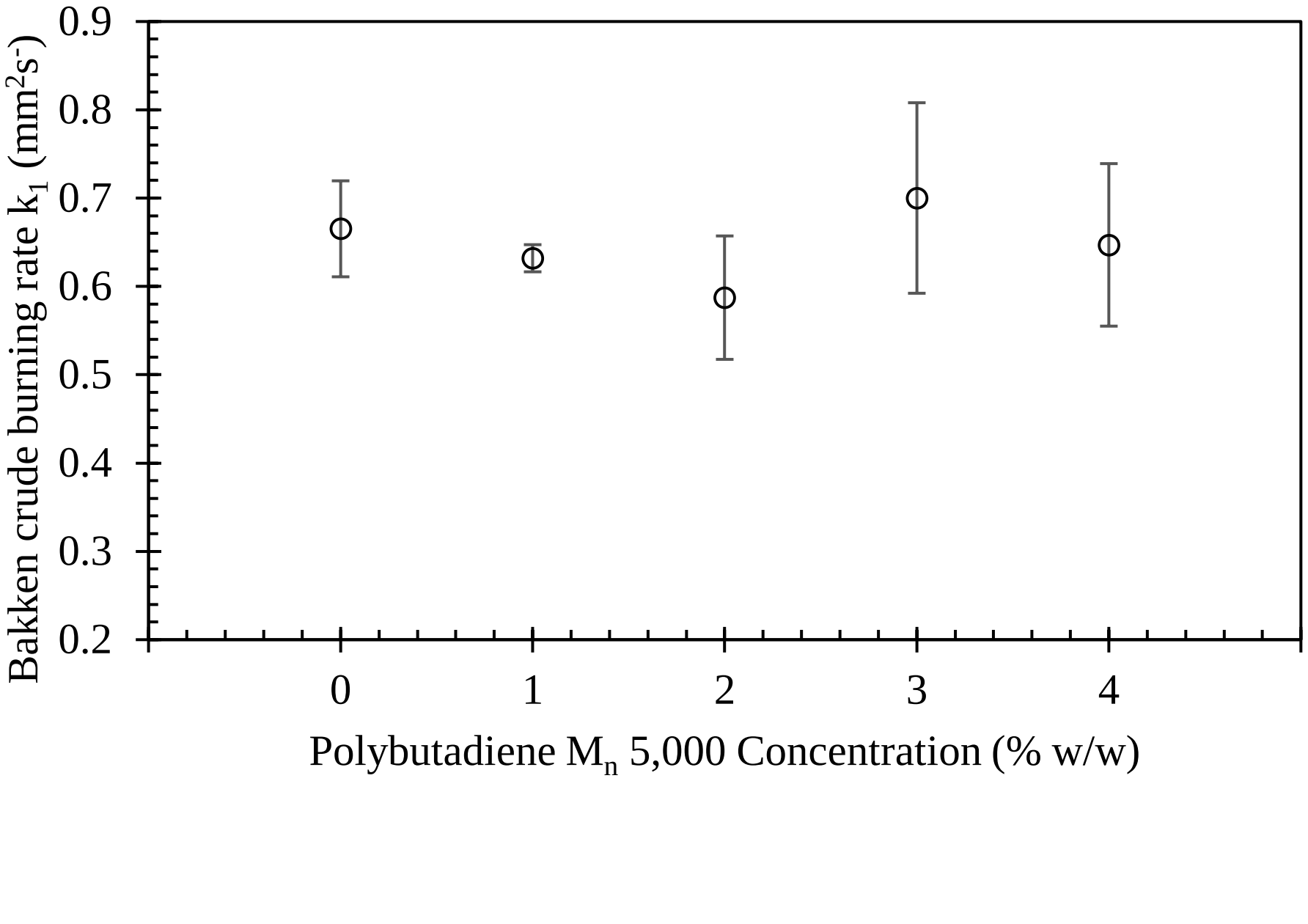

(b)

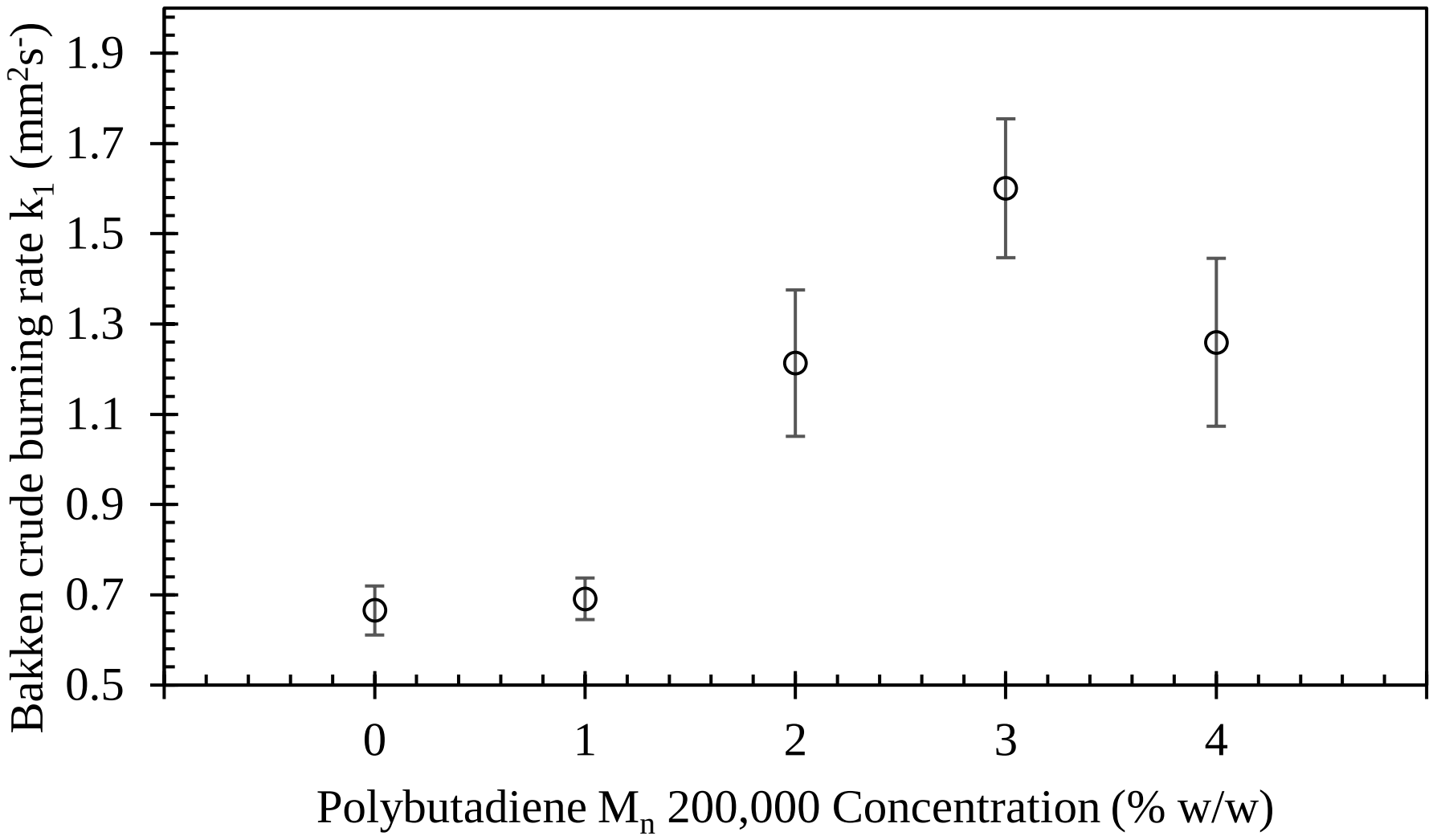

**Figure 7.** Effect of (a) 5k and (b) 200k chain length PBD on Bakken crude oil combustion rate $k_1$. Error bars show the standard deviation of all experiments.

(a)

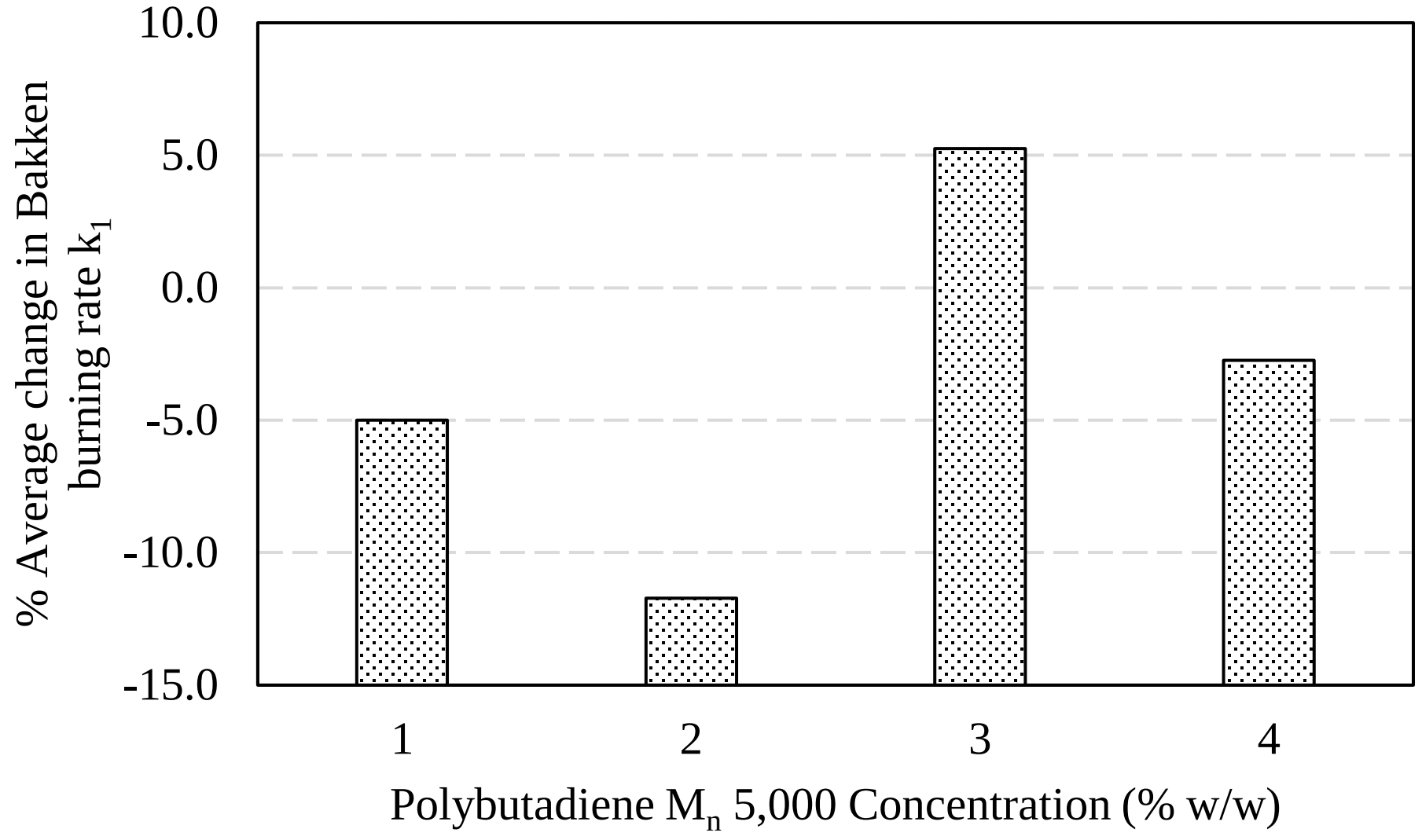

(b)

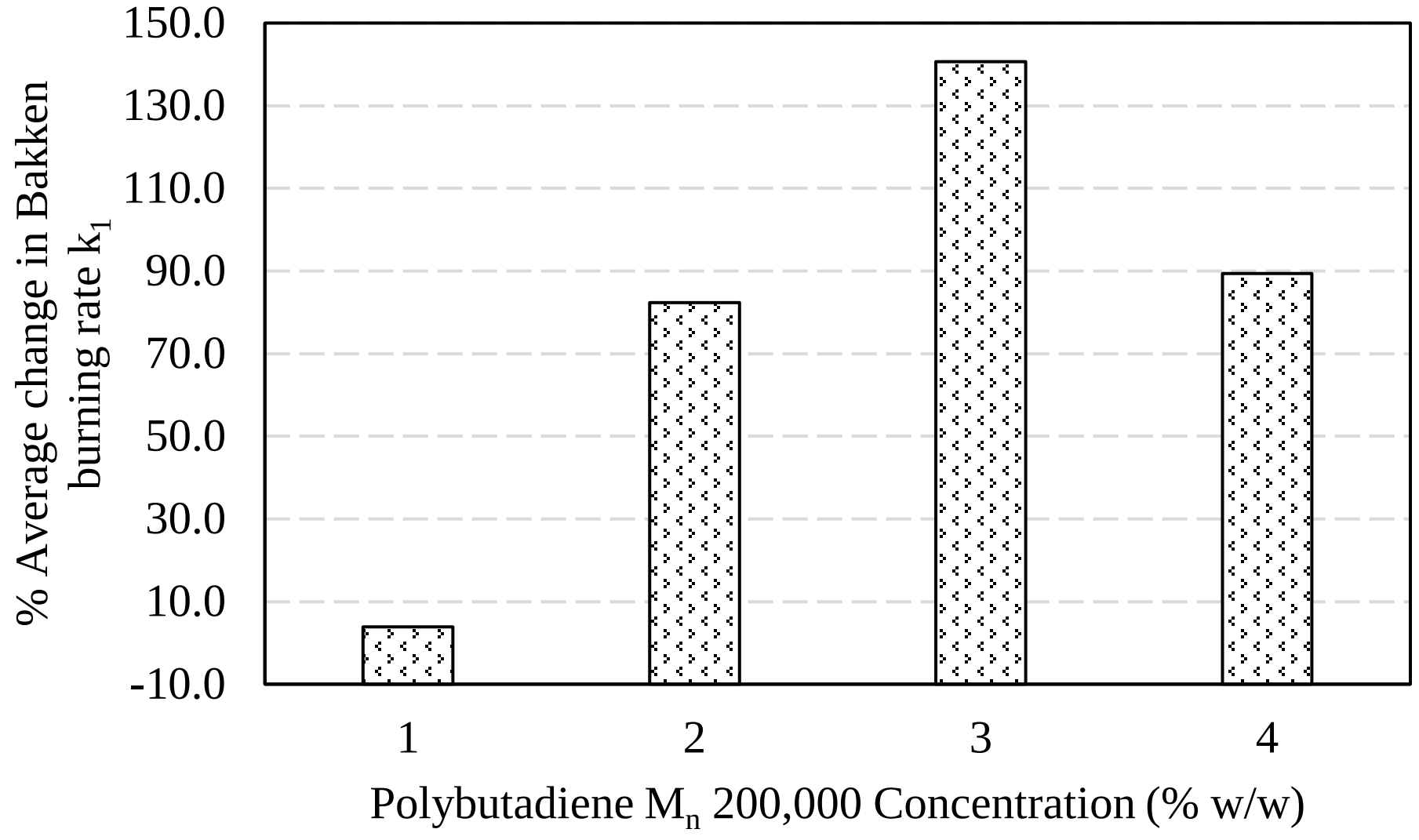

**Figure 8.** Effect of (a) 5k and (b) 200k chain length PBD on Bakken crude oil average combustion rate $k_1$.

As can be seen, PBD5k is much more effective at decreasing combustion rates for both Pennsylvania and Bakken crudes. In comparison, the much-longer length polymer PBD200k is much more effective in increasing combustion rates in both crudes.

Polymer combustion is a complex phenomenon. Subjected to the high heat and radiation of combustion, polymers inside the liquid droplet bulk can be expected to undergo cross-linking, scissoring, and thermal degradation [27]. Thermal degradation of a polymer chain leads to rupture of primary valence bonds, which leads to a lower overall molecular weight and yields smaller components [28]. These smaller components, typically methane and $C_{2-4}$ hydrocarbons [29] frequently burn faster compared to their parent components. As an unsaturated polymer, PBD is more susceptible to a thermo-oxidative degradation (such as in a combustion environment) compared to a saturated polymer [30]. It can be expected that PBD200k, because of its longer chain length and therefore more unsaturation compared to PBD5k chain, is more susceptible to thermo-oxidative degradation. It has also been established before that increasing chain length in polymers leads to an increase in thermal conductivity [31], and increased thermal conductivity has been linked before to increased droplet combustion rates [17]. Also, due to easier C-C bond scission in longer chain polymers compared to short chain polymers, PBD200k is more susceptible to thermal degradation compared to PBD5k [32].

Therefore, many factors such as increased thermal degradation and increased heat conductivity lead to PBD200k increasing overall combustion rates for its blends with either crude oil tested. PBD5k, due to its much shorter chain length, possibly suffers from smaller thermal degradation but is effective at decreasing diffusion of lighter crude ends through the liquid bulk, and decreasing surface evaporation rates from the droplet surface, leading to lower overall combustion rate for its blends with either crude. This is further substantiated in Section 3.4 of this manuscript, where a general decrease in flames stand-off ratio is observed at all PBD5k-Pennsylvania crude blends, which indicate a lowered flame speed resulting from a decrease in diffusion and surface evaporation.

As remarked before, combustion regimes of pure crudes lack a Zone V, which is characterized by a sharp change in combustion rate from the preceding Zone IV. It is also seen that Zone V combustion behavior is quite uniform and shows similarities to that of a pure fuel (**Figure 3**). From this, we can conclude that it is mainly leftover PBD that is burning in Zone V. This is supported by the fact that more PBD is added to a blend, further away from pure crude its burning rate gets (**Figure 9**, **Figure 10**). Furthermore, SEM analysis of the combustion residue shows polymeric residue (Section 3.5) for one of the fuel blends tested (Section 3.5).

**Figure 9** and **Figure 10** show the comparison of combustion rate $k_2$for various fuel blends, with combustion rate of pure fuel provided for reference. For both PBD5k and PBD200k, it is seen that as the weight % of PBD increases the combustion rate $k_2$ also increases. This can be explained by PBD degrading by two distinct events when heated dynamically such as in a combustion environment. The first stage is depolymerization that yields lighter, volatile products. Material that does not depolymerize cyclizes and crosslinks to form a residue that degrades in the second stage. Increasing the amount of PBD being degraded shifts the reaction towards the first stage [33], thereby increasing the $k_2$ combustion rates for blends that have more PBD in them.

(a)

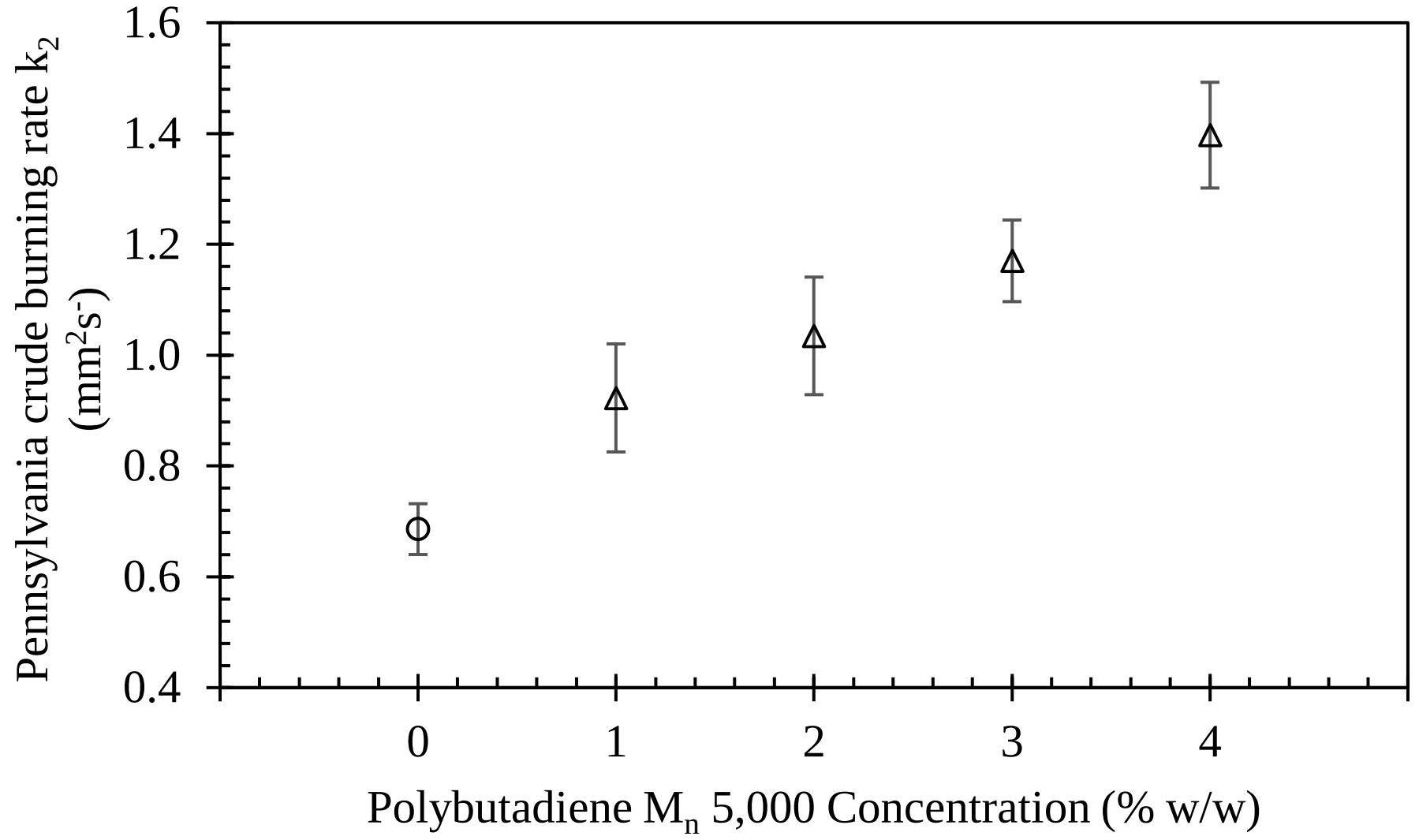

(b)

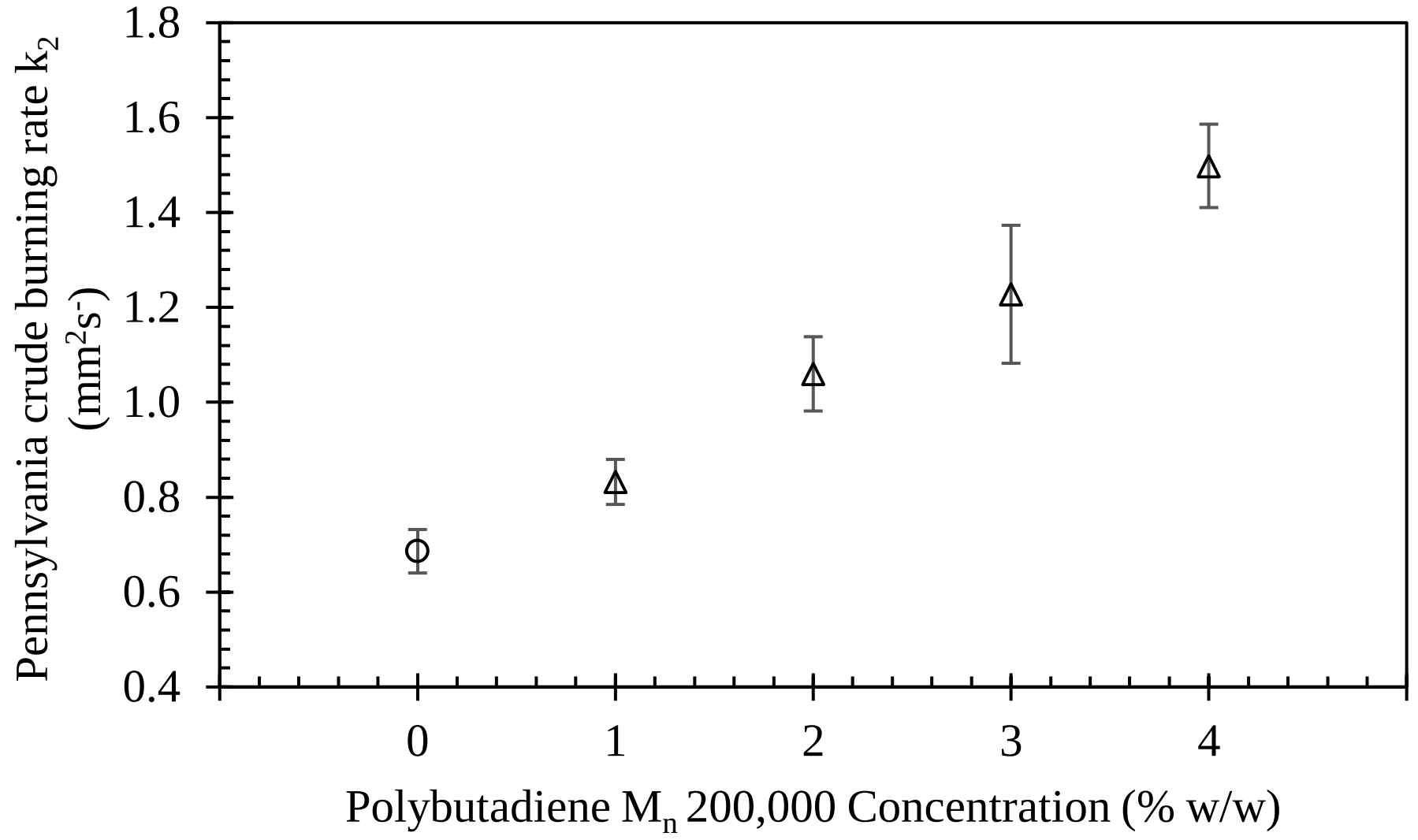

**Figure 9.** Effect of (a) 5k and (b) 200k chain length PBD on Pennsylvania crude oil combustion rate $k_2$(△). Error bars show the standard deviation of all experiments. Combustion rate $k_1$for pure Pennsylvania crude is provided for reference (○).

(a)

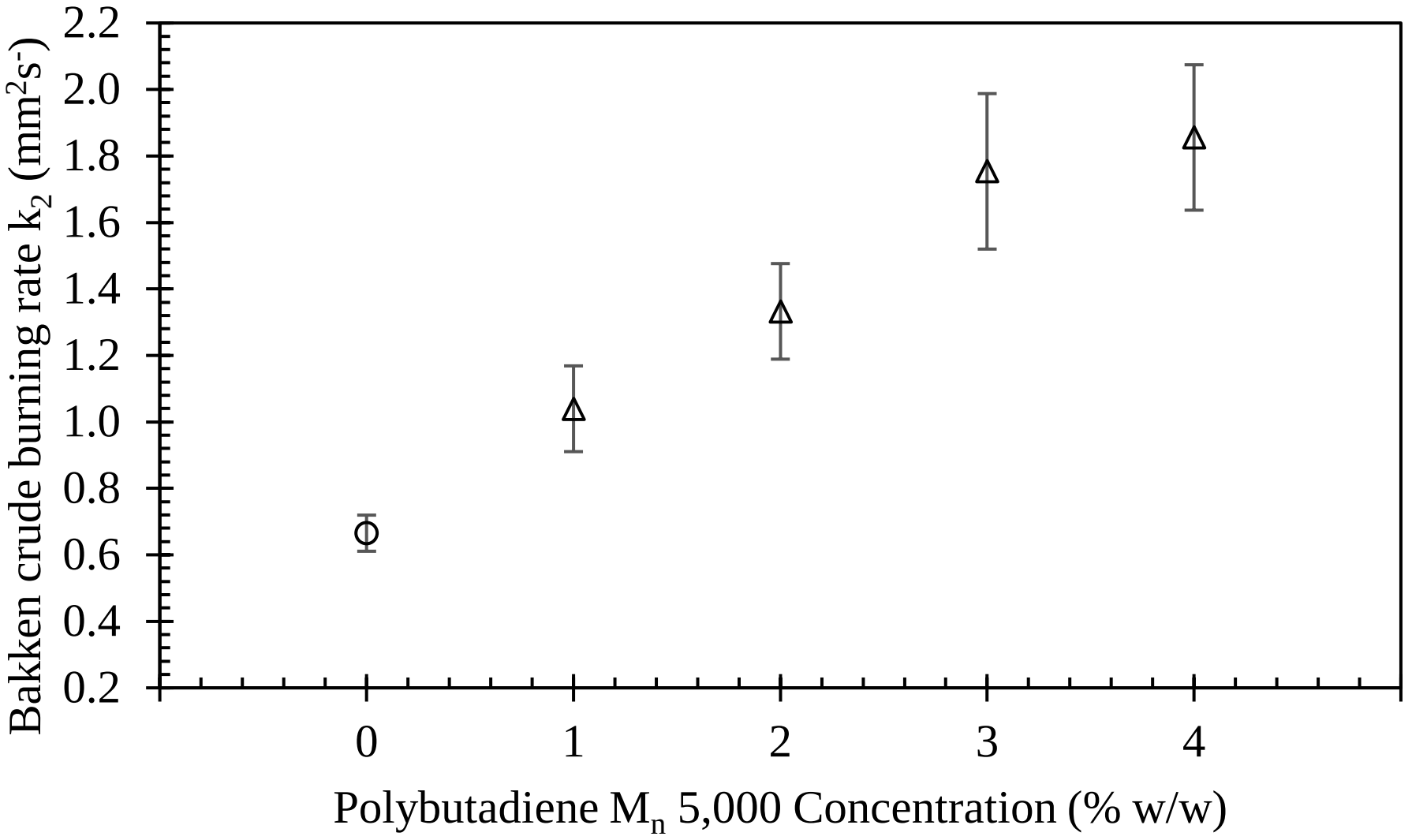

(b)

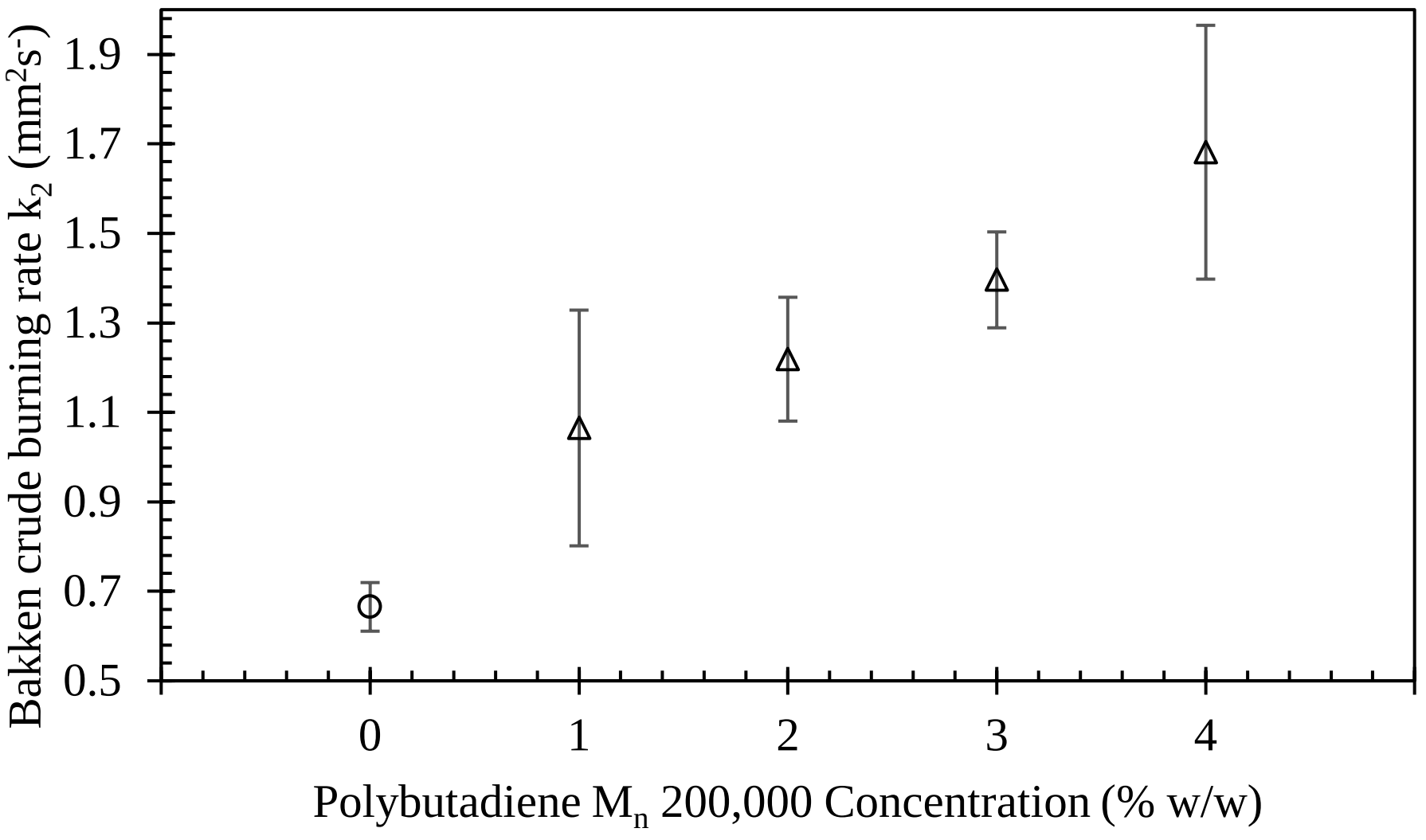

**Figure 10.** Effect of (a) 5k and (b) 200k chain length PBD on Bakken crude oil combustion rate $k_2$ (△). Error bars show the standard deviation of all experiments. Combustion rate $k_1$for pure Bakken crude is provided for reference (○).

3.2 *Ignition delay of fuel droplets*

Many factors influence ignition delay of fuels. There is a physical delay that is associated with the droplet heating to the point of giving off enough vapors to sustain combustion through evaporation, mixing and diffusion of combustible components through the droplet bulk, and a chemical delay [34][35][36][37]. **Figure 11** shows the effect of PBD5k and PBD200k on Pennsylvania crude ignition delay. **Figure 12** shows the change in average ignition delay of Pennsylvania crude oil when PBD5k and PBD200k are added at different concentrations. PBD5k causes a large decrease in ignition delay for Pennsylvania crude at all blend proportions, with the largest decrease of 26% occurring at 1% w/w. Generally, PBD200k causes an increase in ignition delay to Pennsylvania crude, with the largest increases of 6% occurring at 1% and 3% w/w.

PBD5k, a readily combustible liquid, causes a decrease in ignition delay of Pennsylvania crude droplets by dissociating into more combustible smaller components upon heating. Also, possibility of PBD5k dissociation due to chemical action by components in Pennsylvania crude to produce more readily combustible smaller components cannot be ruled out. Because of its larger chain length, a larger ignition delay is associated with PBD200k. However, it does decrease the average vapor pressure of the droplet, causing a net increase in ignition delay for Pennsylvania crude.

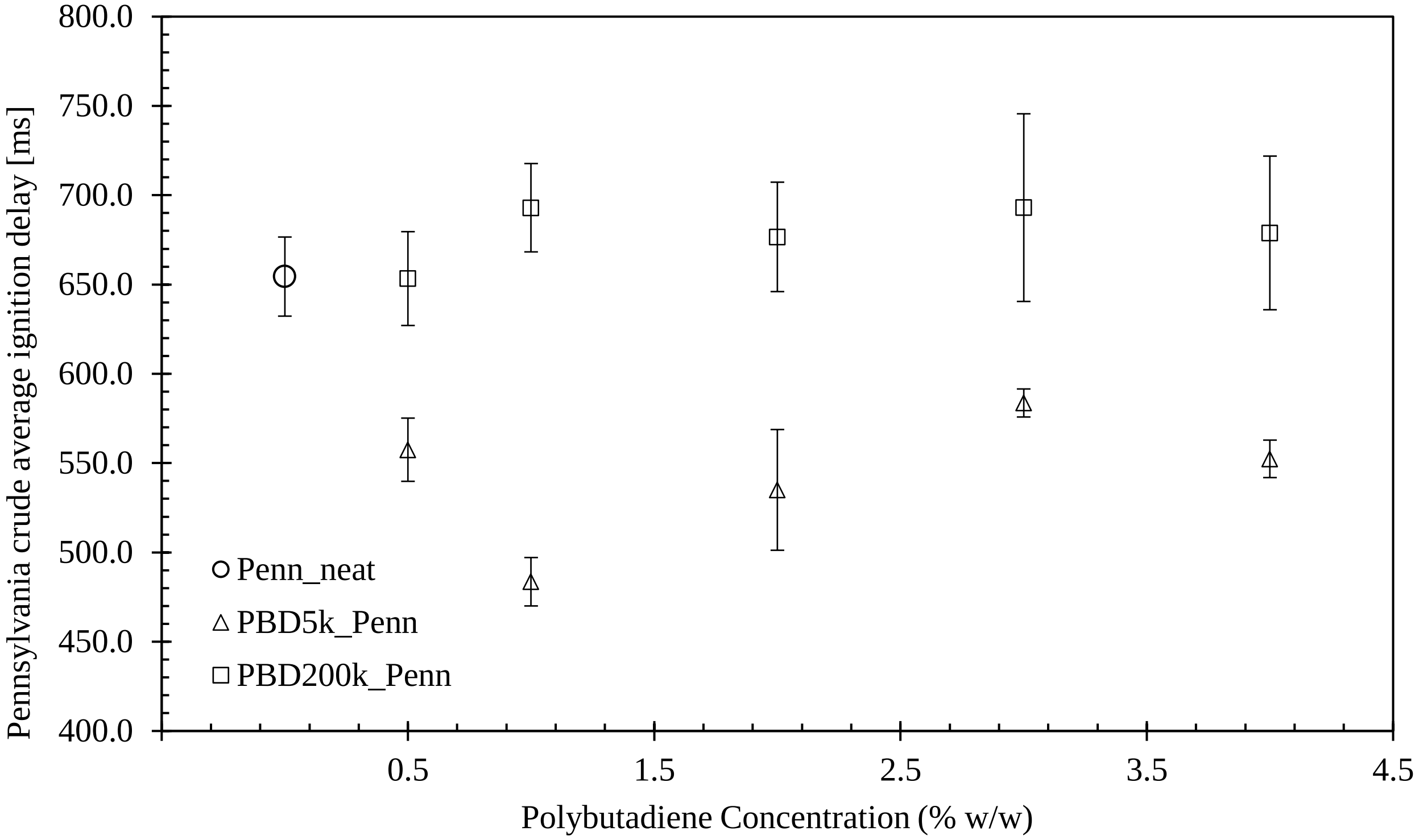

**Figure 11**. Effect of 5k and 200k chain length PBD on ignition delay of Pennsylvania crude

(a)

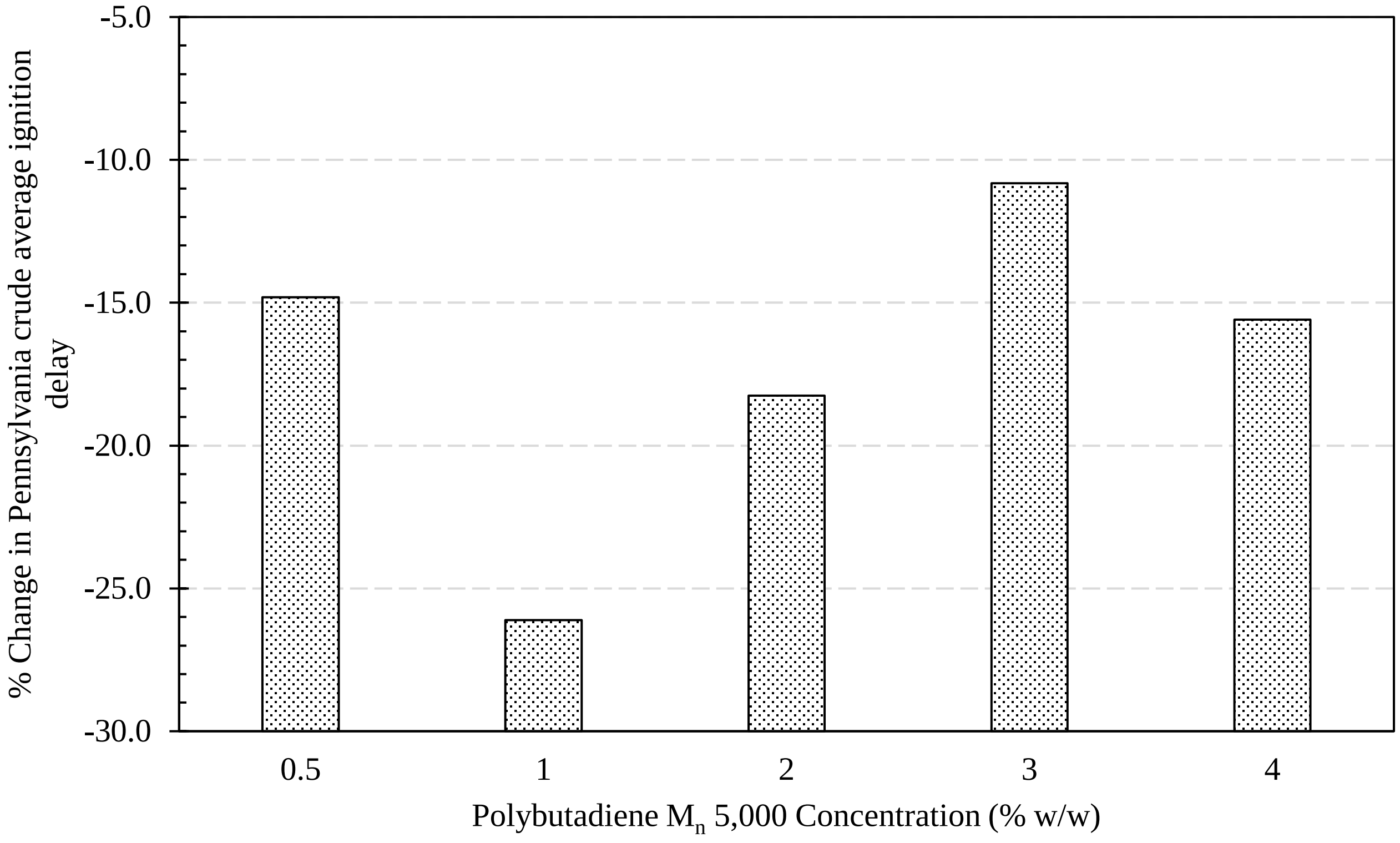

(b)

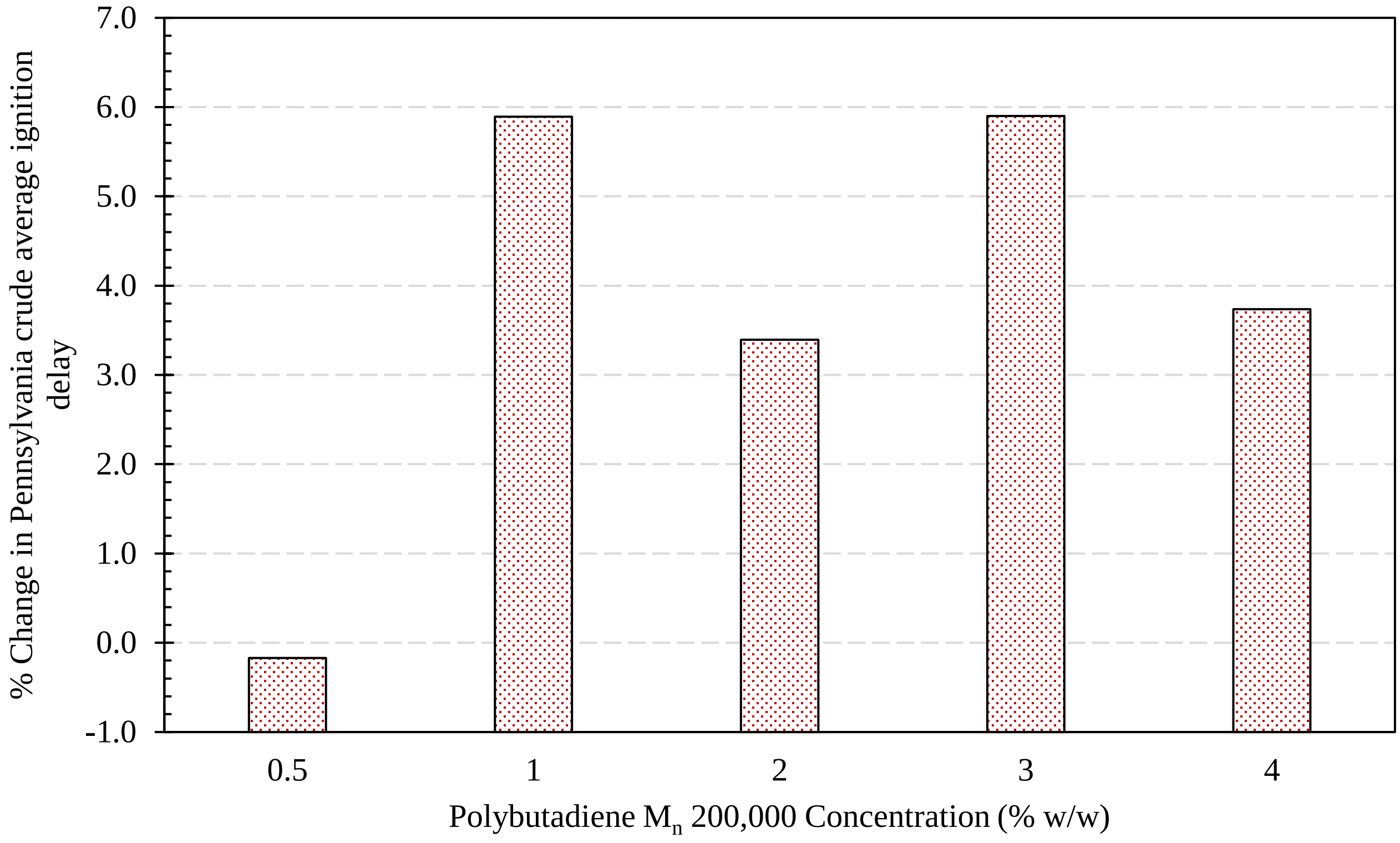

**Figure 12.** Effect of (a) 5k and (b) 200k chain length PBD on average ignition delay of Pennsylvania crude

Both PBD5k and PBD200k cause a large increase in ignition delay for Bakken crude (**Figure 13**). As can be seen from **Figure 14**, a maximum of 52% increase in ignition delay is observed for Bakken crude at addition of 1% PBD5k. A maximum of 42% increase can be observed in ignition delay for Bakken crude at addition of 4% PBD200k. Notably, ignition delay of pure Bakken crude is significantly smaller compared to pure Pennsylvania crude, as noted before by Singh *et al*. [14] because of Bakken crude having more low boiling fractions. Addition of both PBD5k and PBD200k, which are large chain length molecules, decreases vapor pressure in Bakken crude and increases the ignition delay.

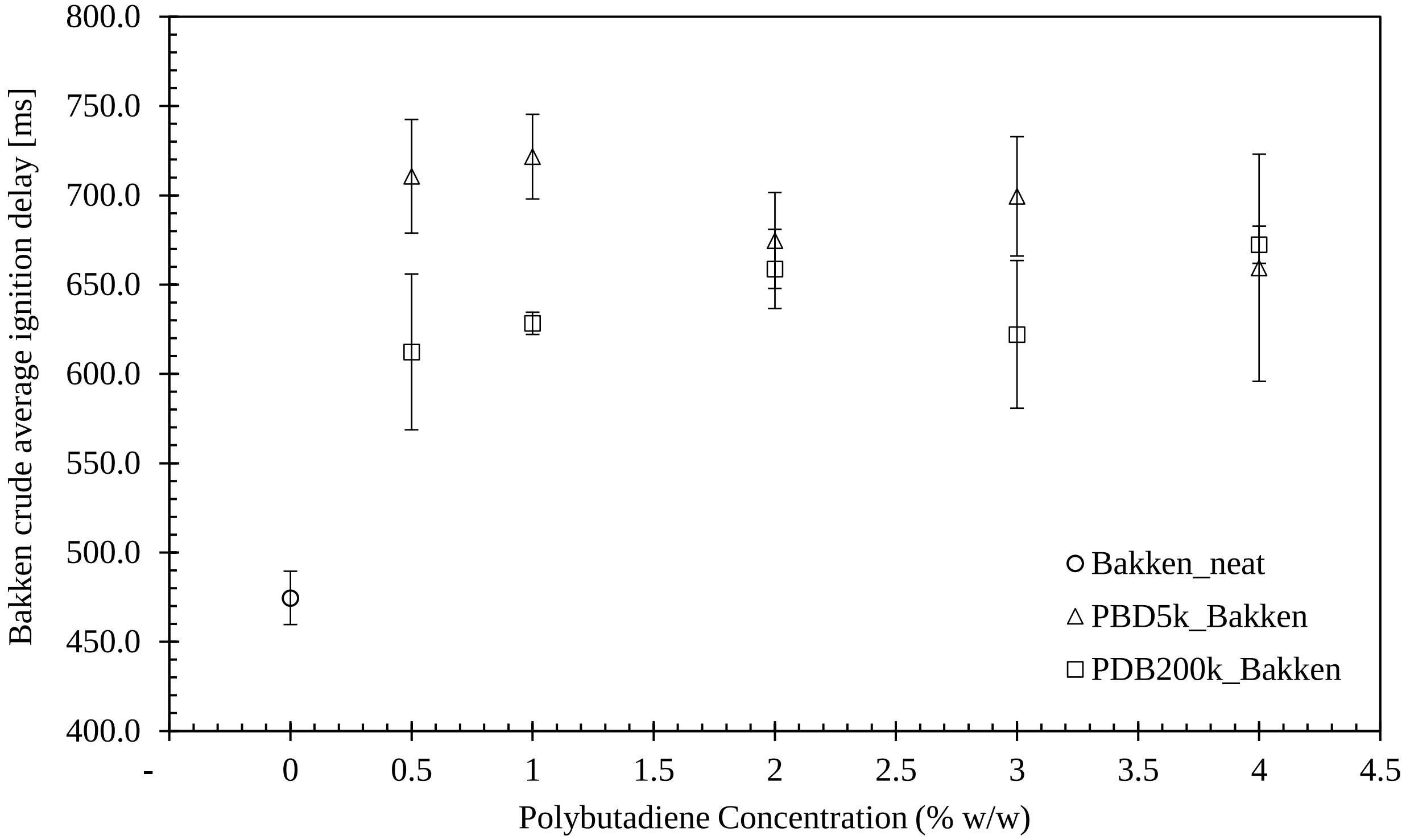

**Figure 13**. Effect of 5k and 200k chain length PBD on ignition delay of Bakken crude

(a)

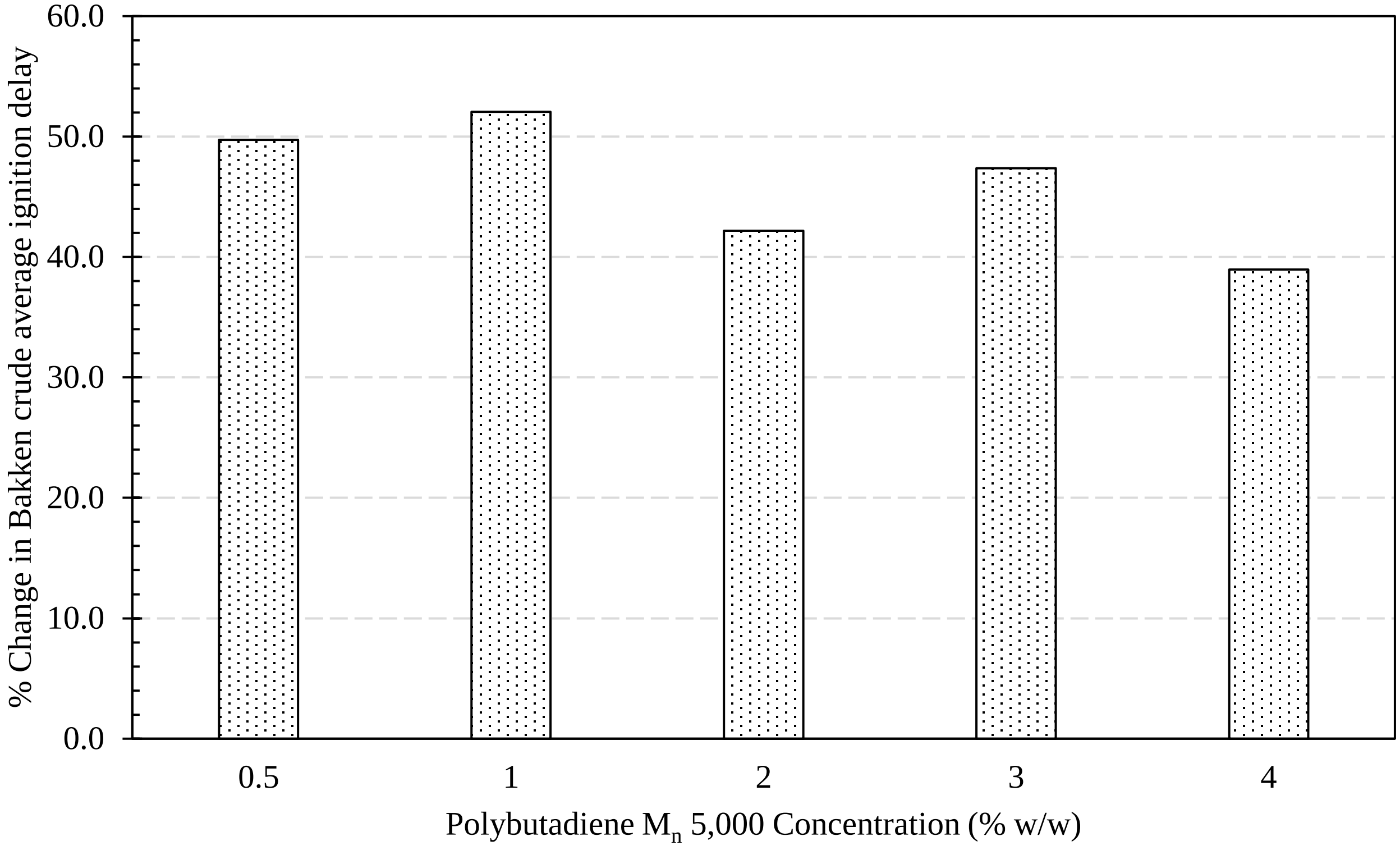

(b)

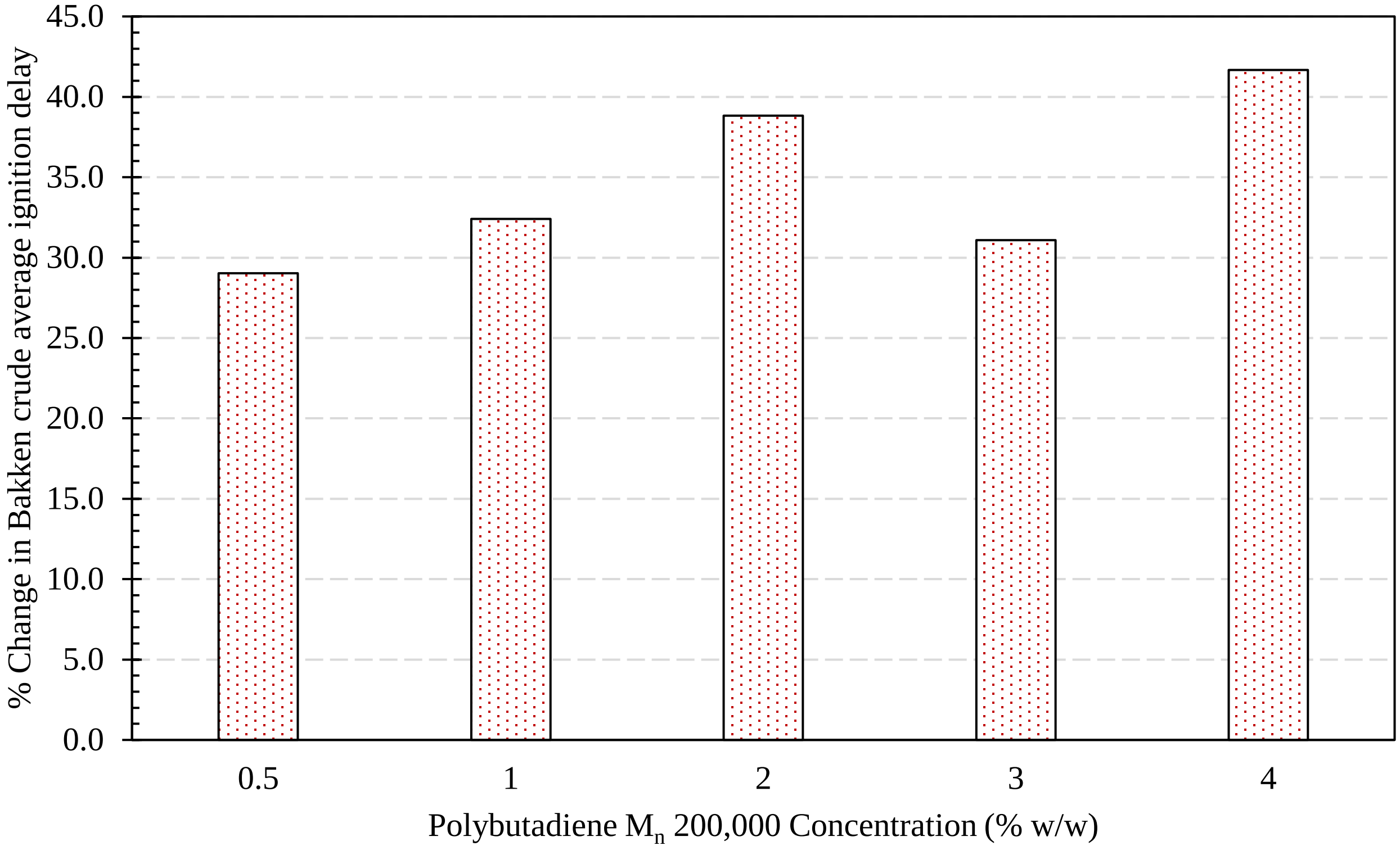

**Figure 14**. Effect of (a) 5k and (b) 200k chain length PBD on average ignition delay of Bakken crude

### 3.3 *Total combustion time of fuel droplets*

Two main factors affect total combustion time. Aside from the obvious droplet combustion rate, microexplosions can cause liquid to escape from the droplet bulk. More intense and more frequent the microexplosions, less the total combustion time. **Figure 15** shows the effect of PBD5k and PBD200k on total combustion time of Pennsylvania crude. **Figure 16** shows the average change in total combustion time of Pennsylvania crude when PBD5k and PBD200k are added.

Generally, PBD5k causes an increase in Pennsylvania crude total combustion time, mainly because combustion rates decline when PBD5k is added to it. The greatest increase of 3.4% is seen at 3% PBD5k. Similarly, an increase in combustion rates as well as increase in microexplosions (the latter is especially pronounced at lager PBD200k concentrations) on addition of PBD200k generally causes total combustion to decline. The greatest average total combustion time decrease of 15% is seen at 4% PBD200k.

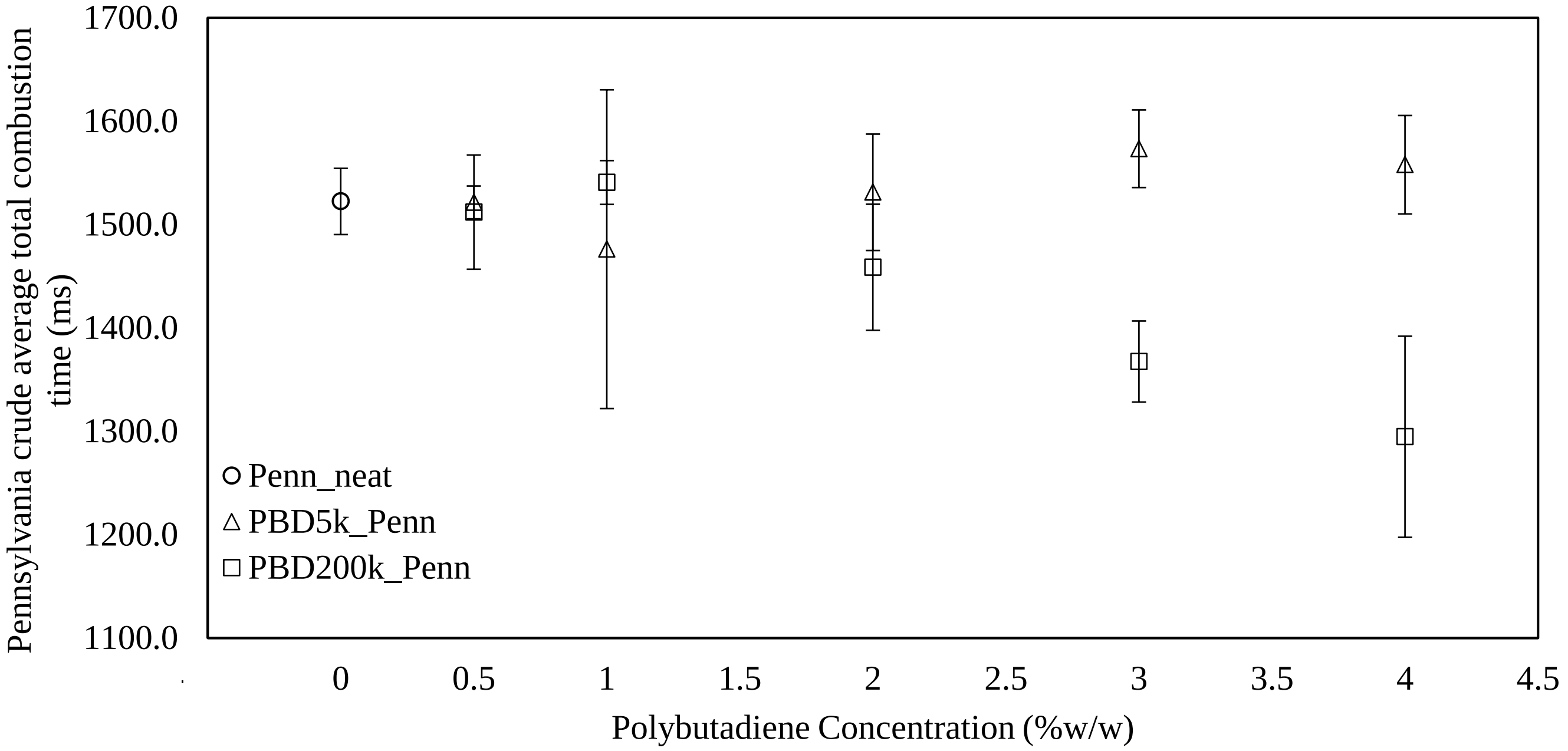

**Figure 15**. Effect of 5k and 200k chain length PBD on total combustion time of Pennsylvania crude

(a)

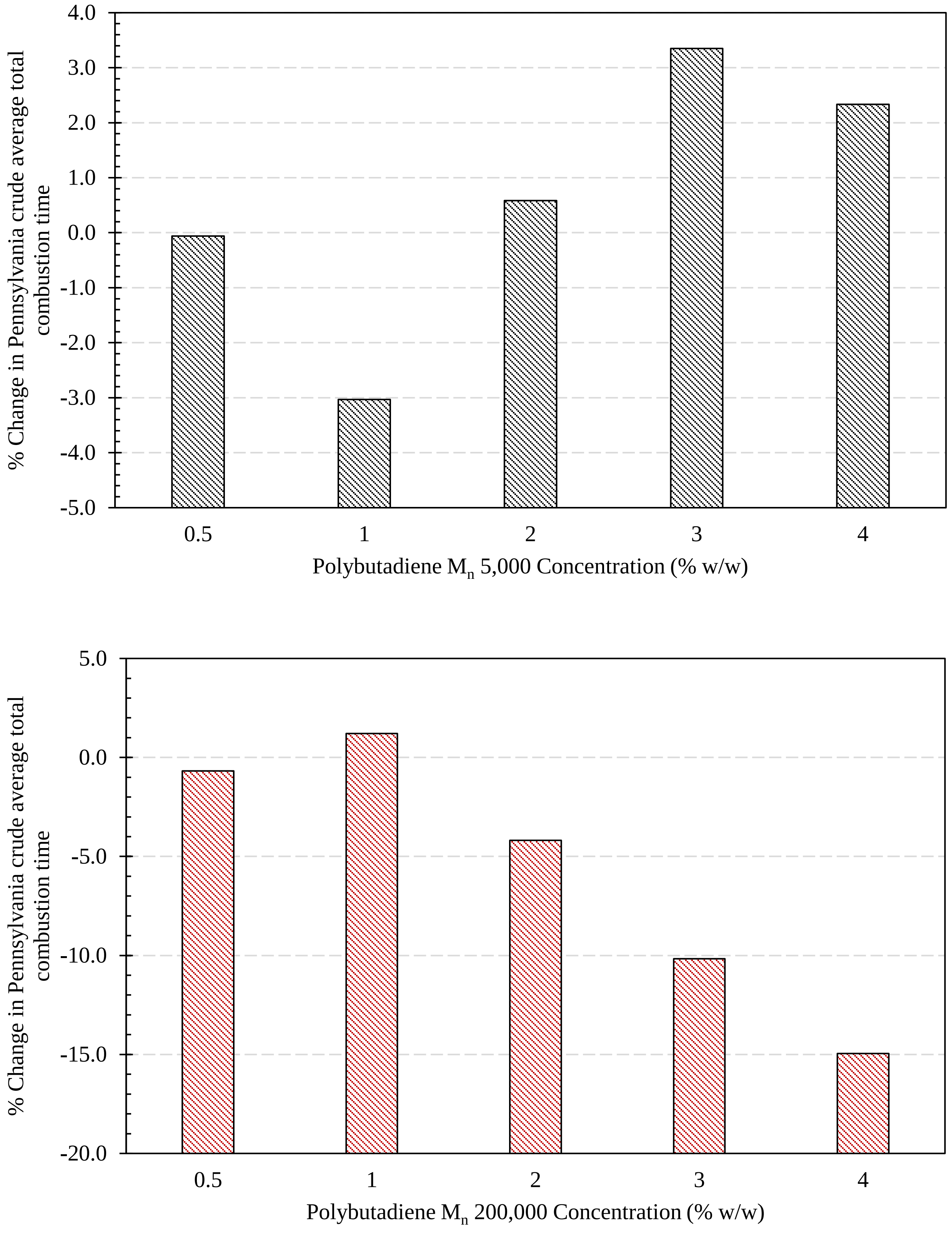

**Figure 16.** Effect of (a) 5k and (b) 200k chain length PBD on average total combustion time of Pennsylvania crude

**Figure 17** shows the change in total combustion time of Bakken crude at various PBD5k and PBD200k polymer concentrations. **Figure 18** shows the average change in total combustion time when PBD5k and PBD200k are added at different proportions. For PBD5k, the general trend is a decrease in total combustion time, despite the decrease in combustion rates. This can be explained by the general increase in microexplosion intensity. The greatest decrease of 8% is seen at 3% PBD5k. Adding PBD200k to Bakken crude greatly decreases total combustion time because of large increase in microexplosion intensity (especially at large PBD200k concentrations) and a large increase in combustion rates. The largest average total combustion time decrease of 28% is noted at 4% PBD200k.

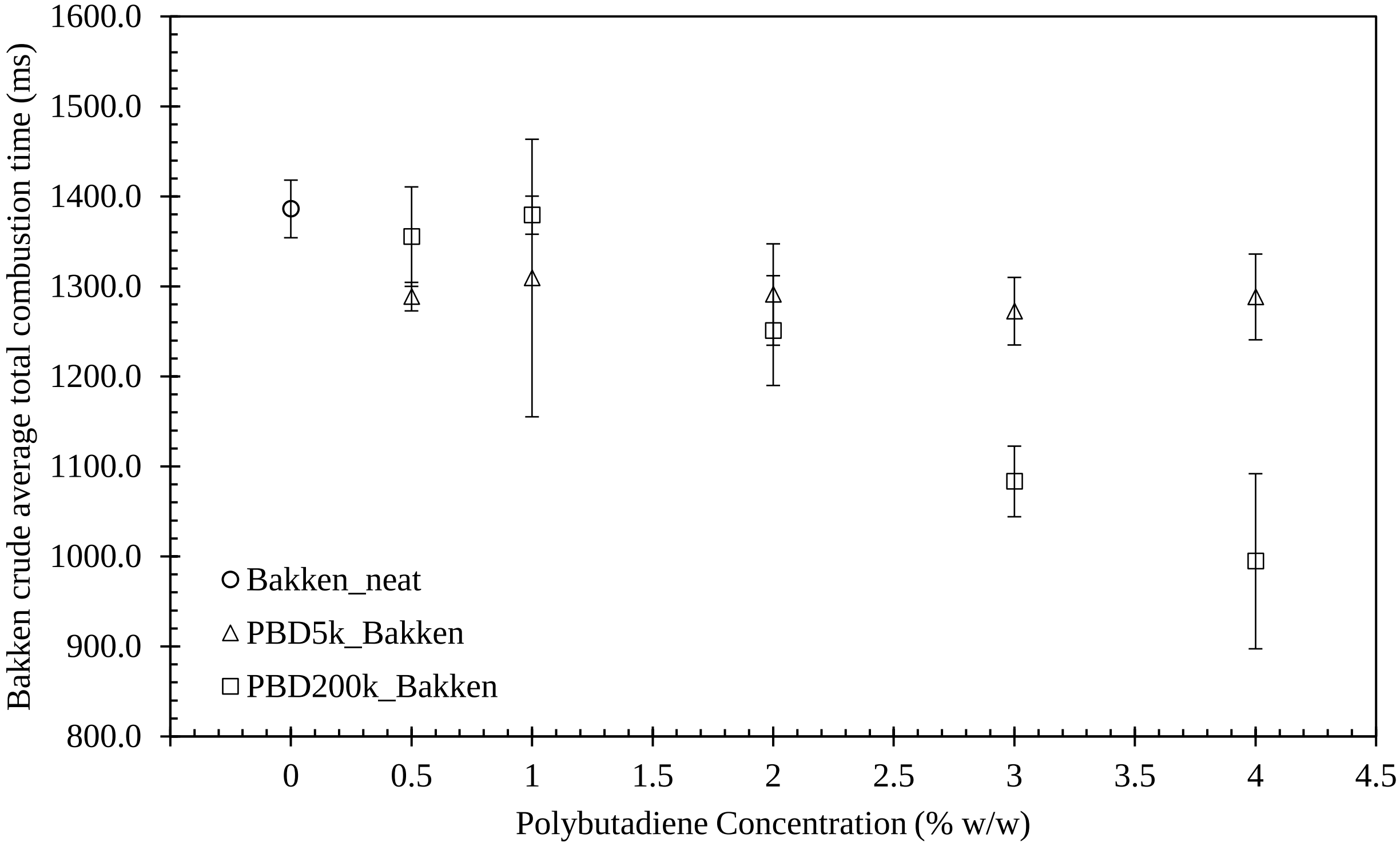

**Figure 17**. Effect of 5k and 200k chain length PBD on total combustion time of Bakken crude

(a)

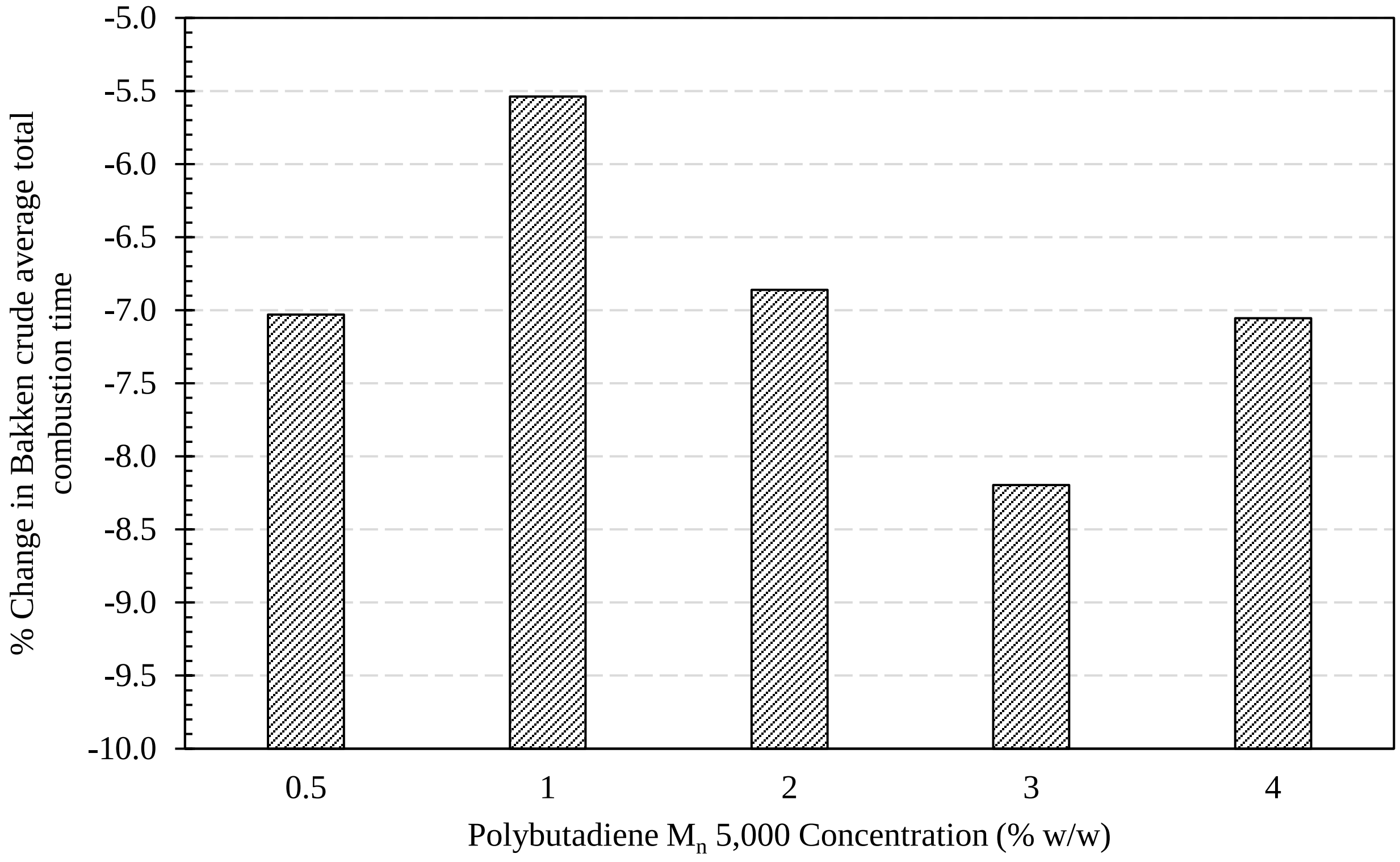

(b)

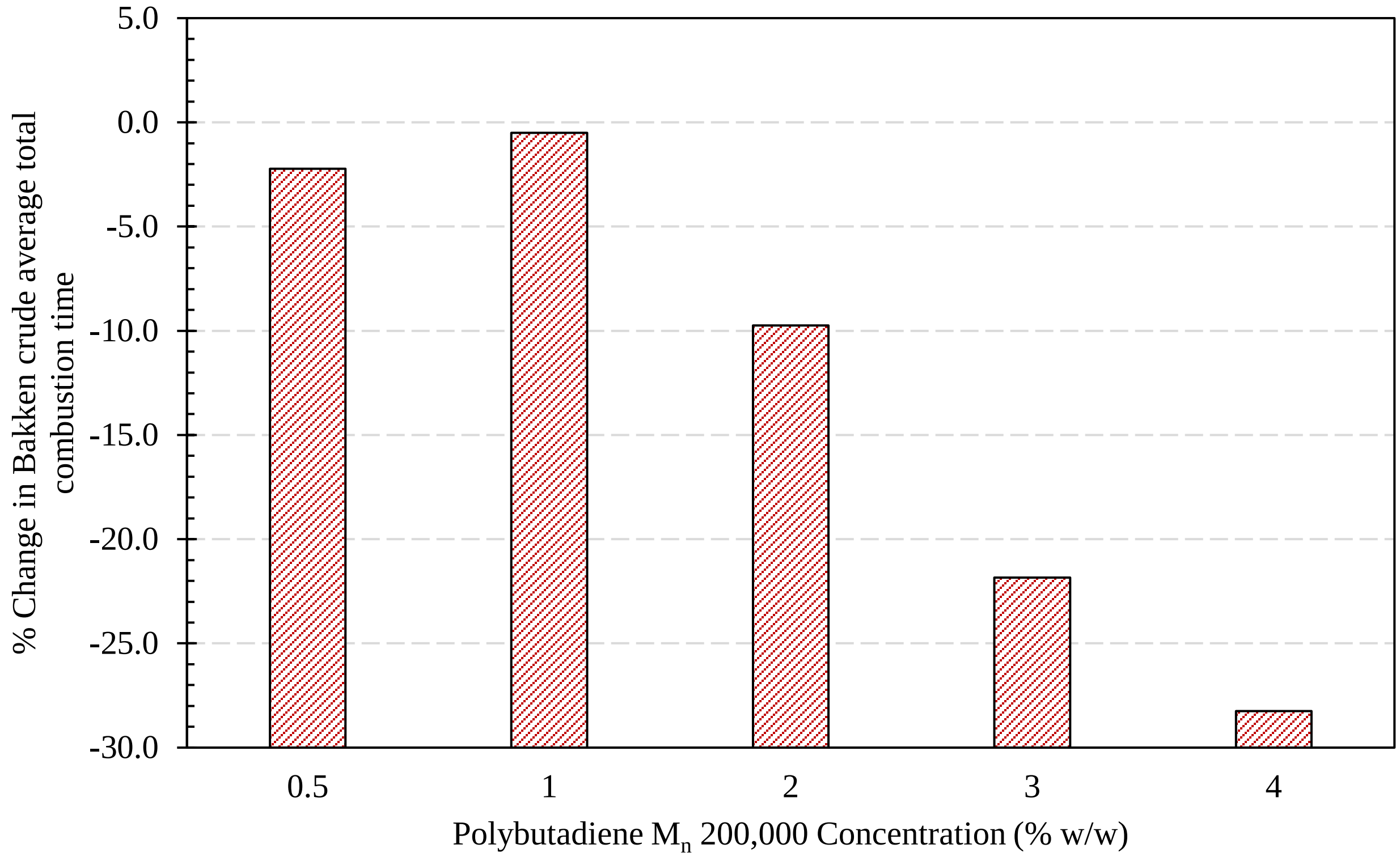

**Figure 18.** Effect of (a) 5k and (b) 200k chain length PBD on average total combustion time of Bakken crude

3.4 *Flame stand-off ratio (FSR) of fuel droplets*

FSR is defined as the ratio of instantaneous flame diameter $d_f$ to instantaneous droplet diameter $d_d$. It can reveal important physical insight into the processes at work. Two major phenomena control FSR: thermophoretic flux and Stefan flux. Thermophoretic flux is caused by the temperature gradient in the combustion zone and causes the flame to move towards the surface. Stefan flux is caused by the outgassing of the vapor from the droplet surface and causes the flame to move away from the surface. Therefore, a fuel with a smaller vapor pressure would have a smaller FSR compared to a fuel with a larger vapor pressure.

**Figure 19** shows the image of a 3% PBD5k-Pennsylvania crude blend in its later stages of combustion as captured by the CCD camera. **Figure 20** shows the FSR for pure Pennsylvania crude vs its blends with 0.5%, 1%, 2%, 3%, and 4% PBD5k. This blend was chosen because of the significant effect of PBD5k on Pennsylvania crude on reducing its combustion rate as observed in Section 3.1 of this manuscript.

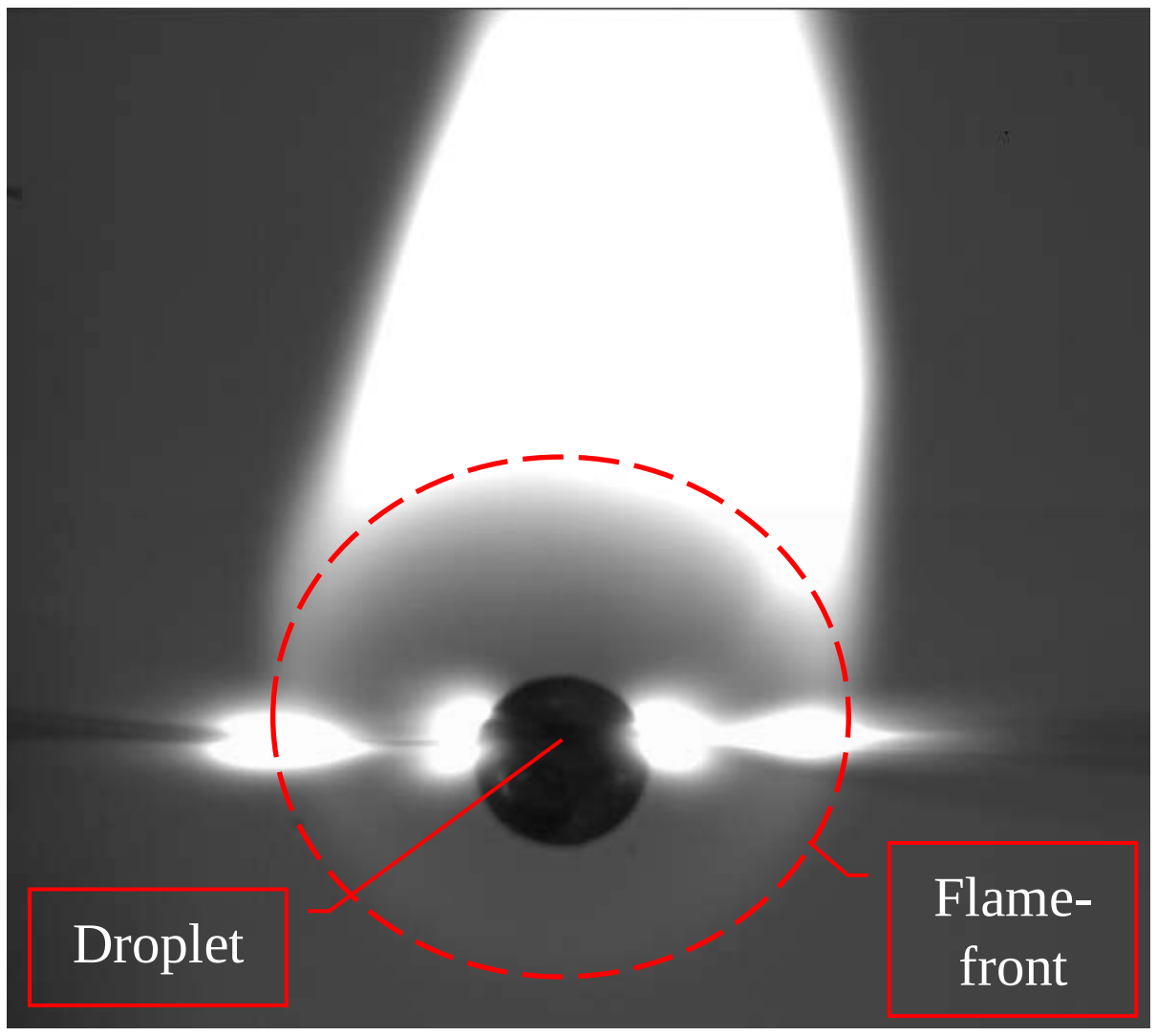

**Figure 19**. Flame structure of a 3% PBD5k-Pennsylvania crude fuel blend droplet undergoing combustion

In general, the FSR for all PBD5k blends with Pennsylvania crude is smaller compared to pure Pennsylvania crude. This points to a general reduction in Stefan flux during crude droplet combustion when PBD5k is added. As noted before, PBD5k is a large-chain, viscous compound which when added to a light hydrocarbon mixture such as Pennsylvania crude can lead to a general decrease in vapor pressure and therefore general decrease in Stefan flux.

(a)

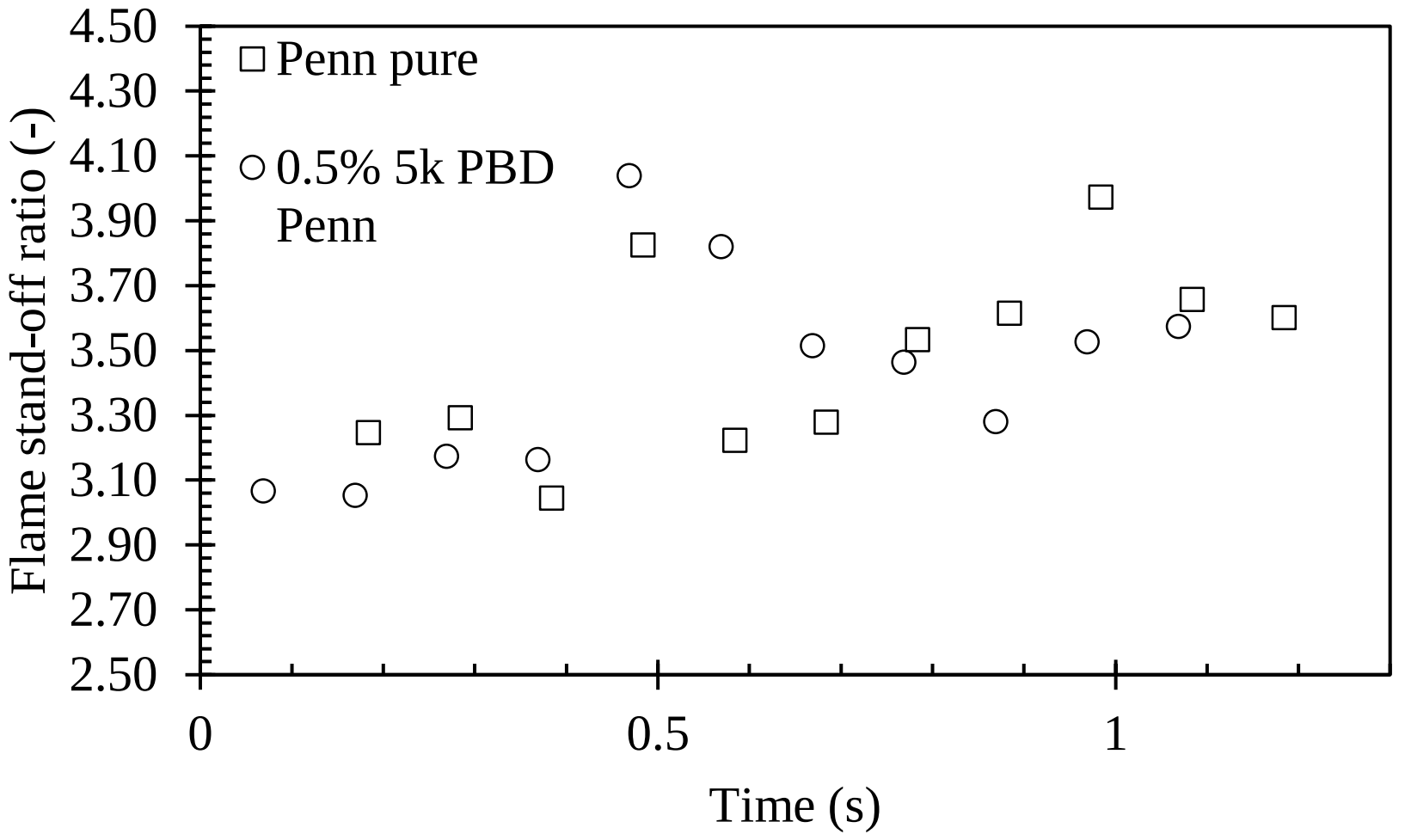

(b)

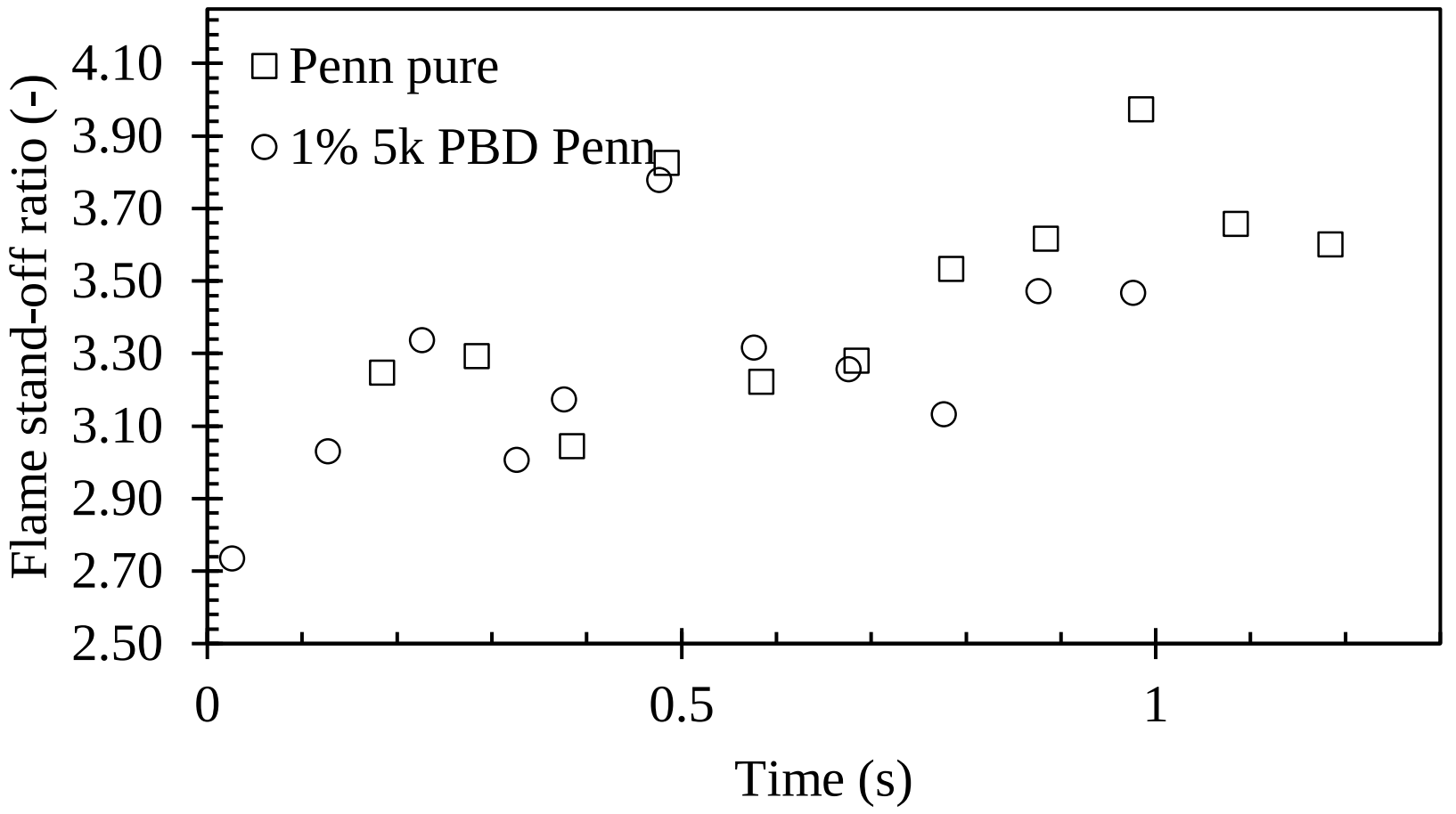

(c)

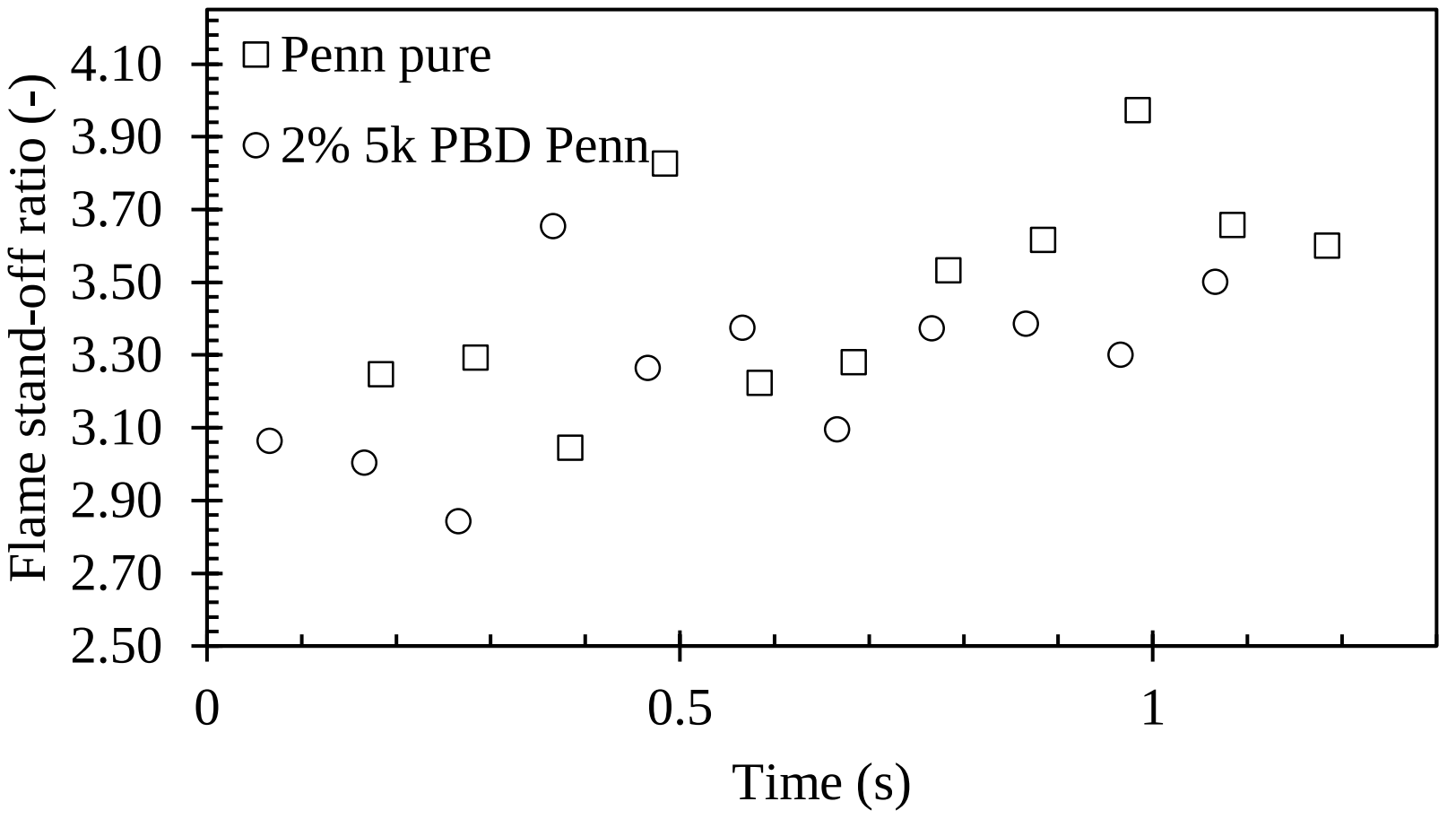

(d)

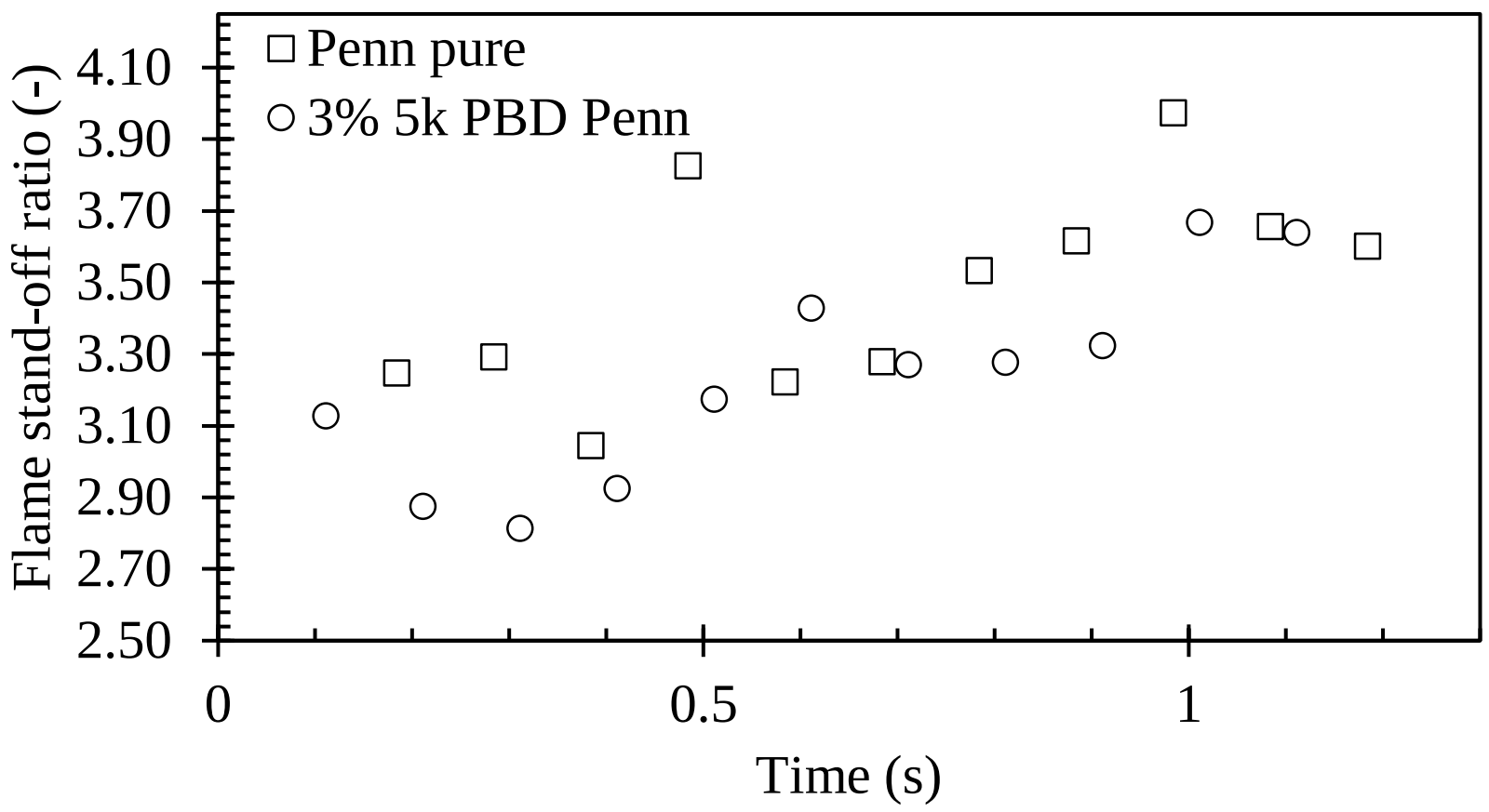

(e)

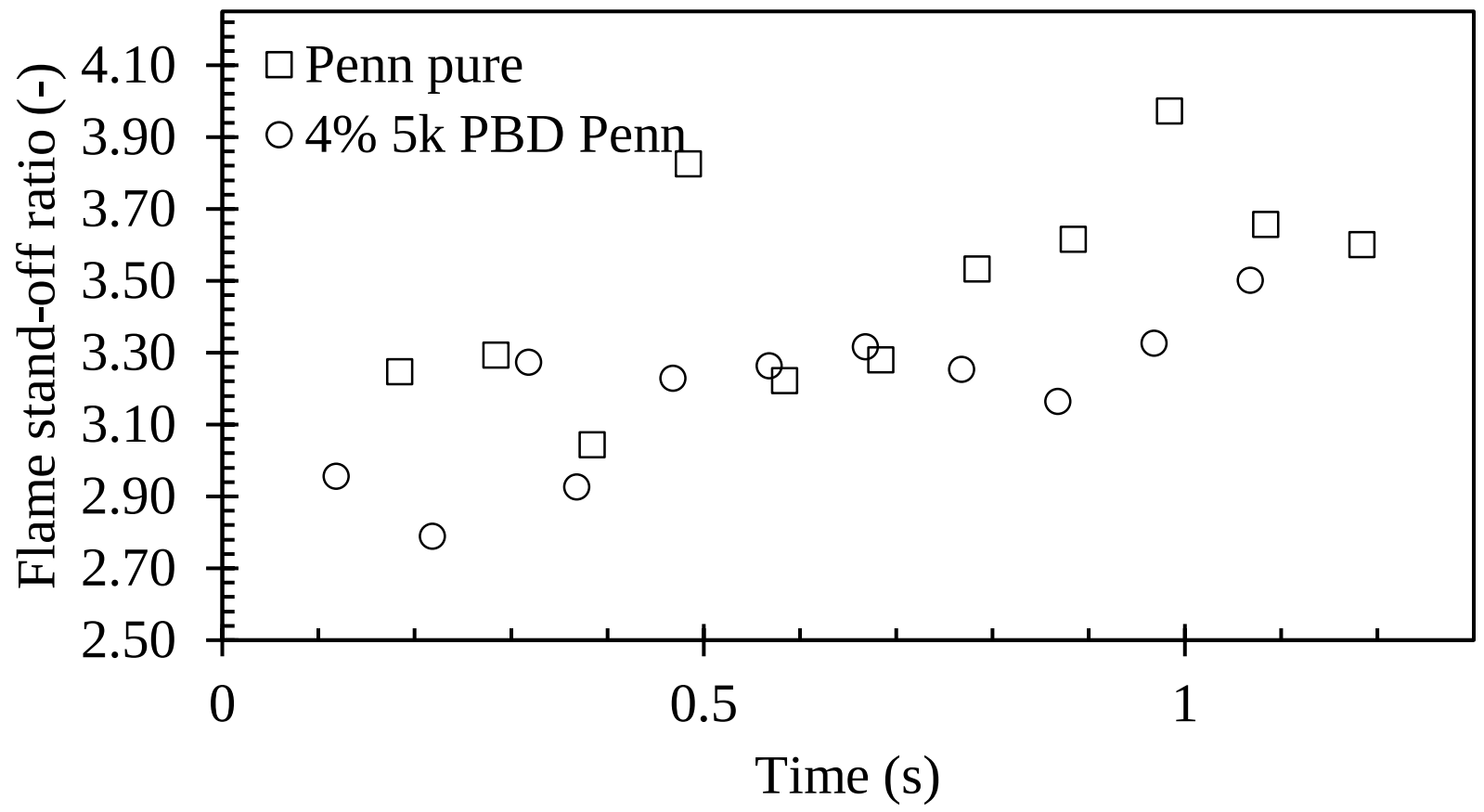

**Figure 20.** Comparison of flame stand-off ratios (FSR) for Pennsylvania crude oil at (a) 0.5%, (b) 1%, (c) 2%, (d) 3%, and (e) 4% w/w polymer blends

### 3.5 *Soot residue analysis*

Soot residue can reveal important information about the combustion process, and determination of particle size can assist in selecting appropriate respiratory personal protection equipment for fire-fighting parties that respond to crude oil fires. **Figure 19** shows the flame structure of a fuel blend droplet undergoing combustion, where soot incandescence is visible in the flame. All crudes and crude blends tested in this work left a soot residue on the thin supporting fibers, and scanning electron microscope (SEM) analysis of several pure crudes has been presented before by the authors [14]. The same equipment (Hitachi S-4800 SEM) located at University of Iowa Central Microscopy Research Facility was used to test soot residue samples for this work. Due to limited equipment availability, only soot residue from 1%, 2%, 3%, 4% PBD5k blend with Bakken crude as a representative polymer-crude blend was analyzed.

To take an SEM image, the fiber with the residue is mounted on a specialized tape with high electrical conductivity. The soot is gently tapped with a fine-tipped needle to break the soot deposit (**Figure 21**) so the soot structure at the surface and in the bulk of the soot can be analyzed. The sample is then loaded into the SEM machine and images at various magnifications are taken.

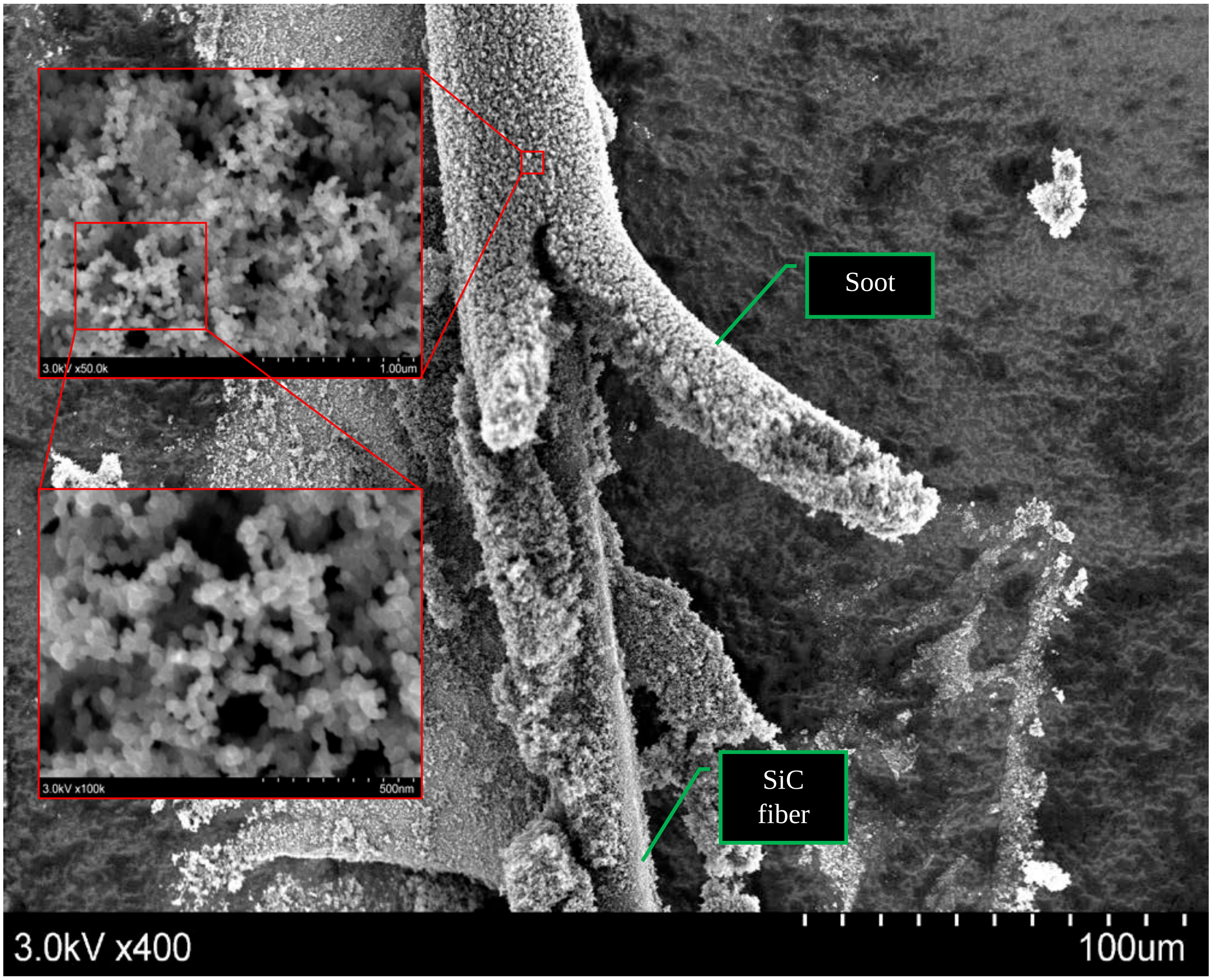

**Figure 21.** SEM images of 1% PBD5k-Bakken blend soot residue, showing soot structure at various levels of magnification. Individual soot particles can be seen

It was reported before that pure Bakken crude leaves behind a loose and spongy soot structure [14] with average particle size ~70µm. However, for all PBD5k-crude blends the structure of the soot is observed to be closely packed, with average particle size ~40µm (**Figure 22**). Additionally, imaging of the soot structure surface reveals globular polymeric structure embedded in the soot residue (**Figure 23**) in one of the samples. This is unburned polymeric residue from the polymer-only combustion in Zone V as discussed in Section 3 of this manuscript. Ultimately, soot structures and individual soot particle sizes are dependent on the chemical make-up of the fuel or fuel blend.

(a)

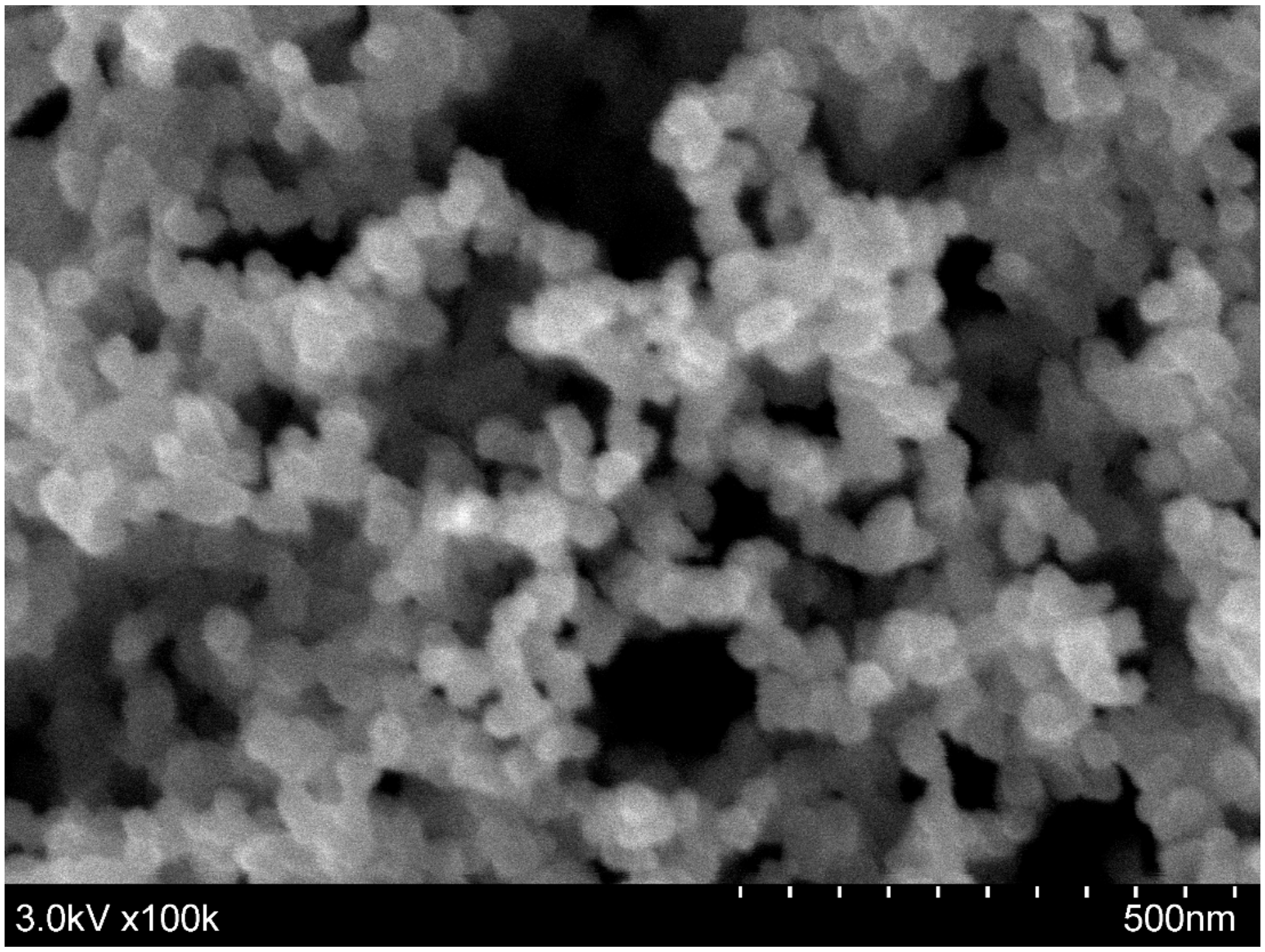

(b)

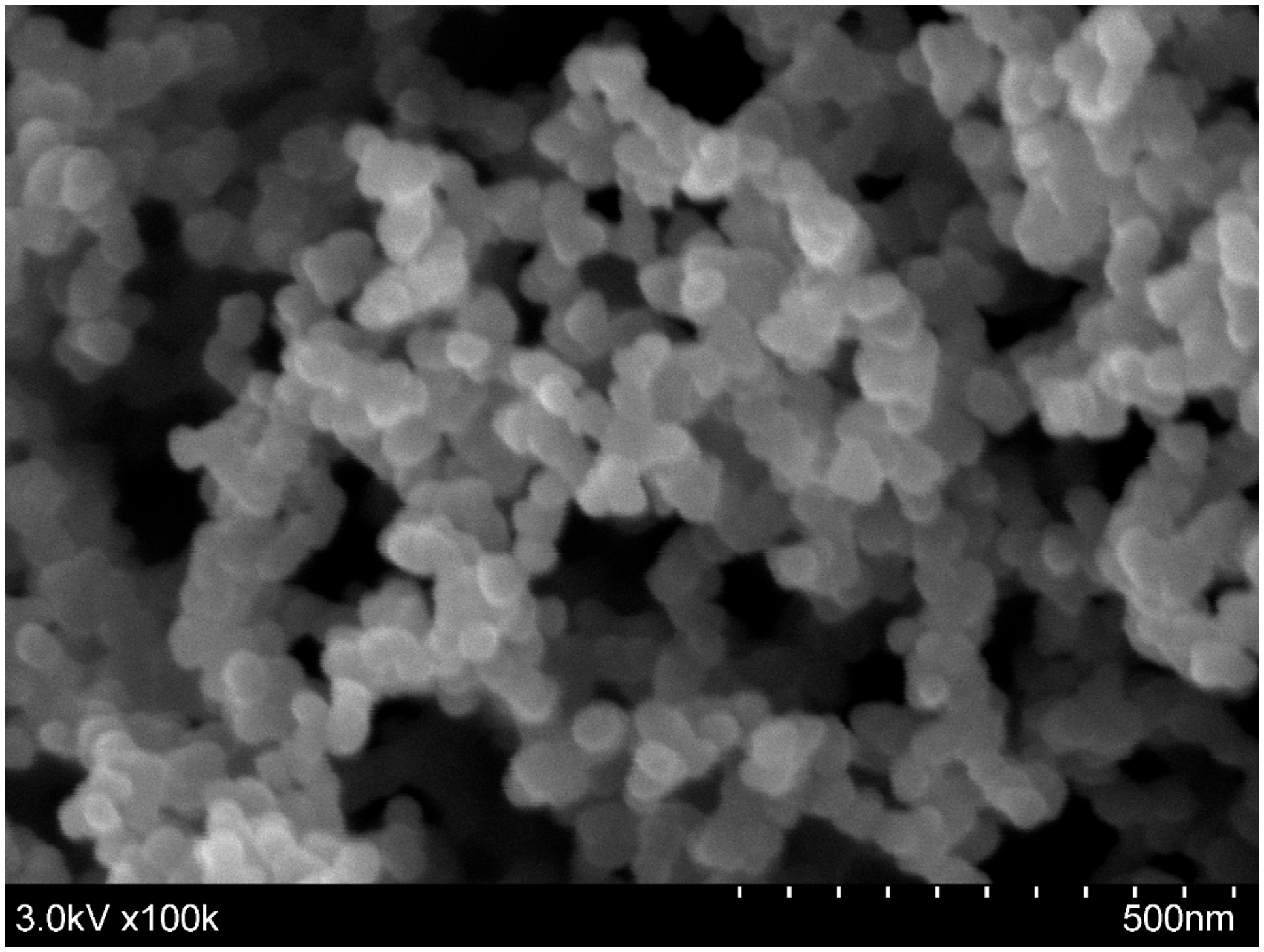

(c)

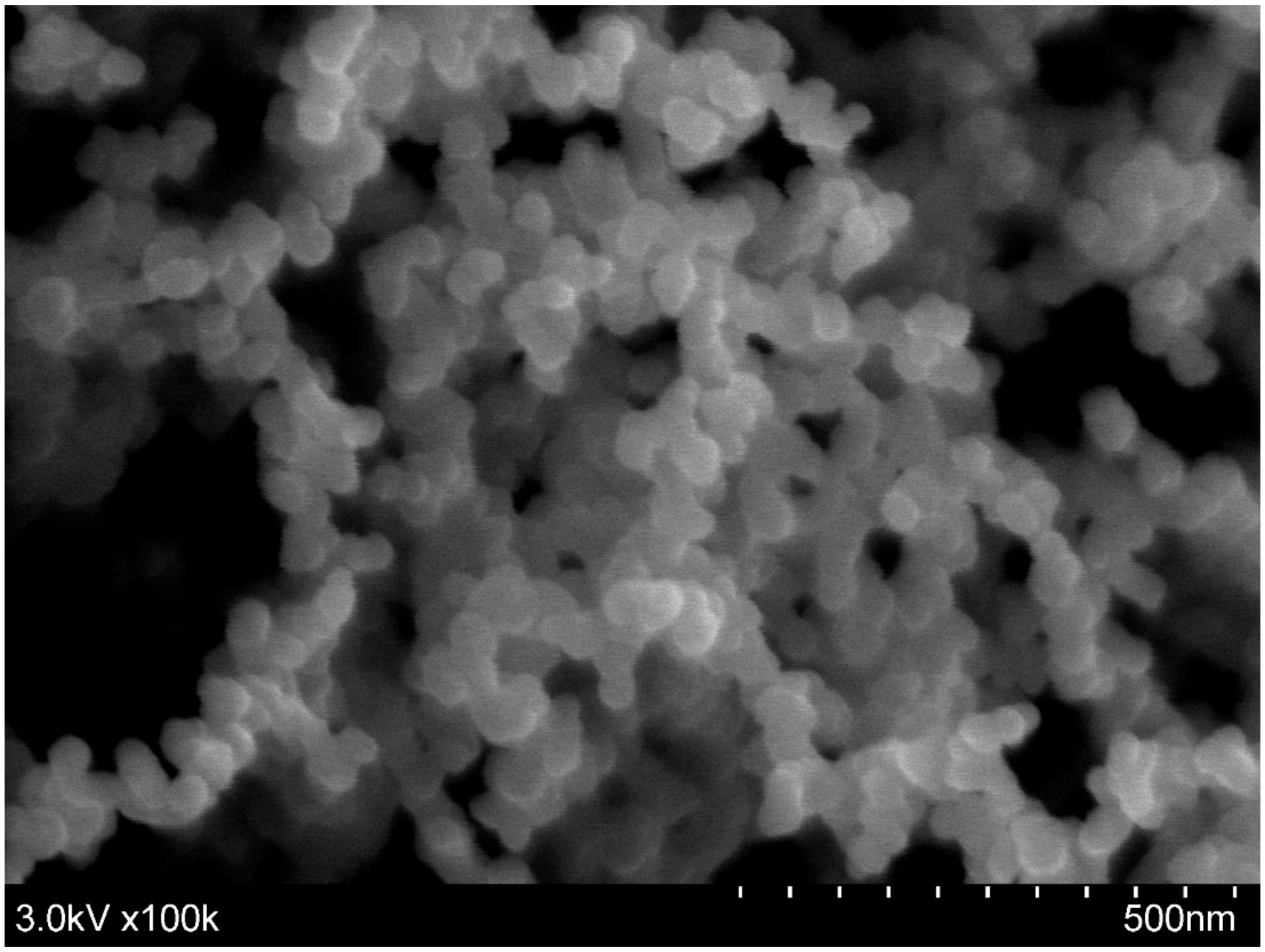

(d)

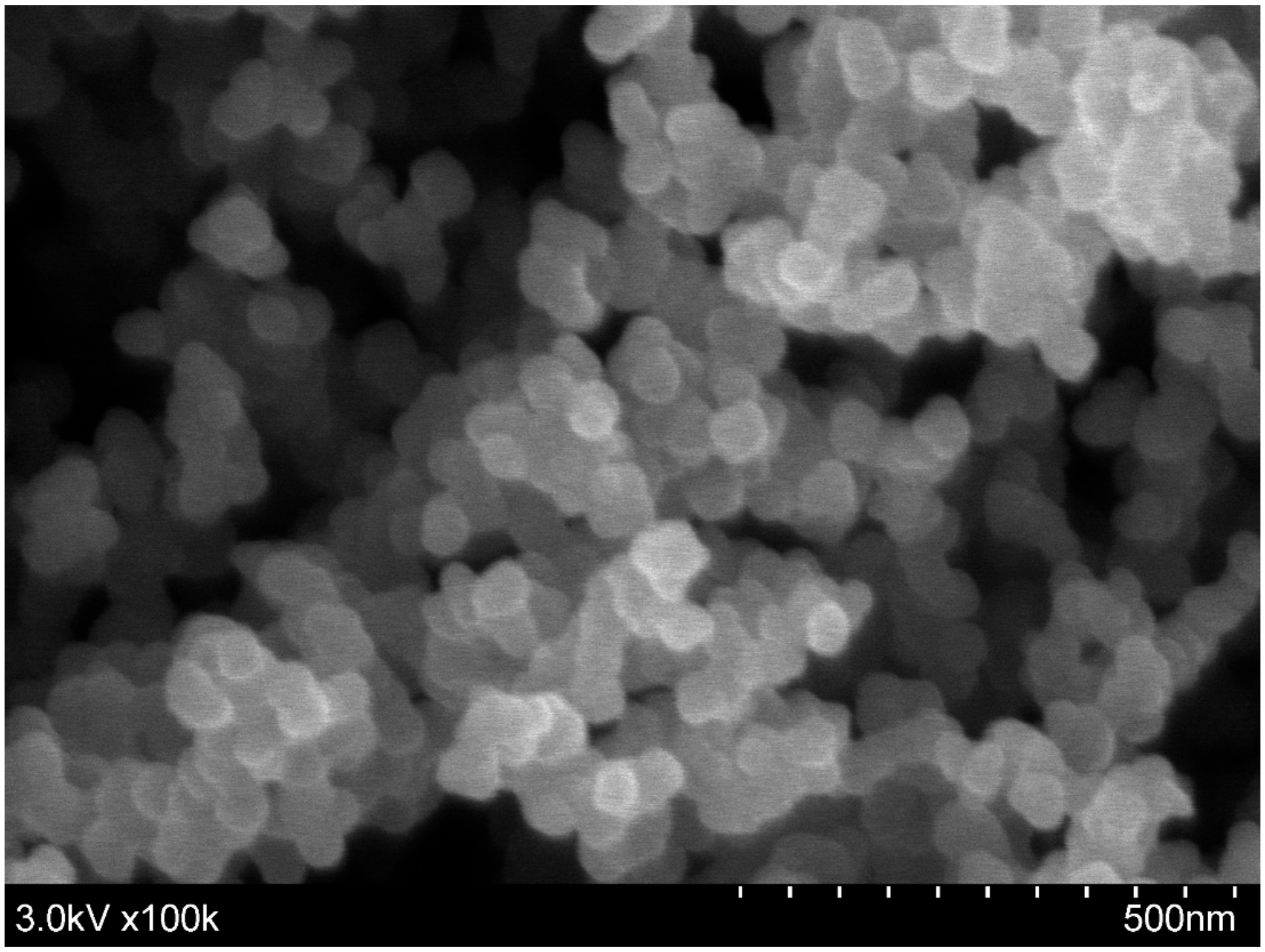

**Figure 22.** Soot particle size comparison for (a) 1%, (b) 2%, (c) 3%, and (d) 4% PBD5k-Bakken crude blends

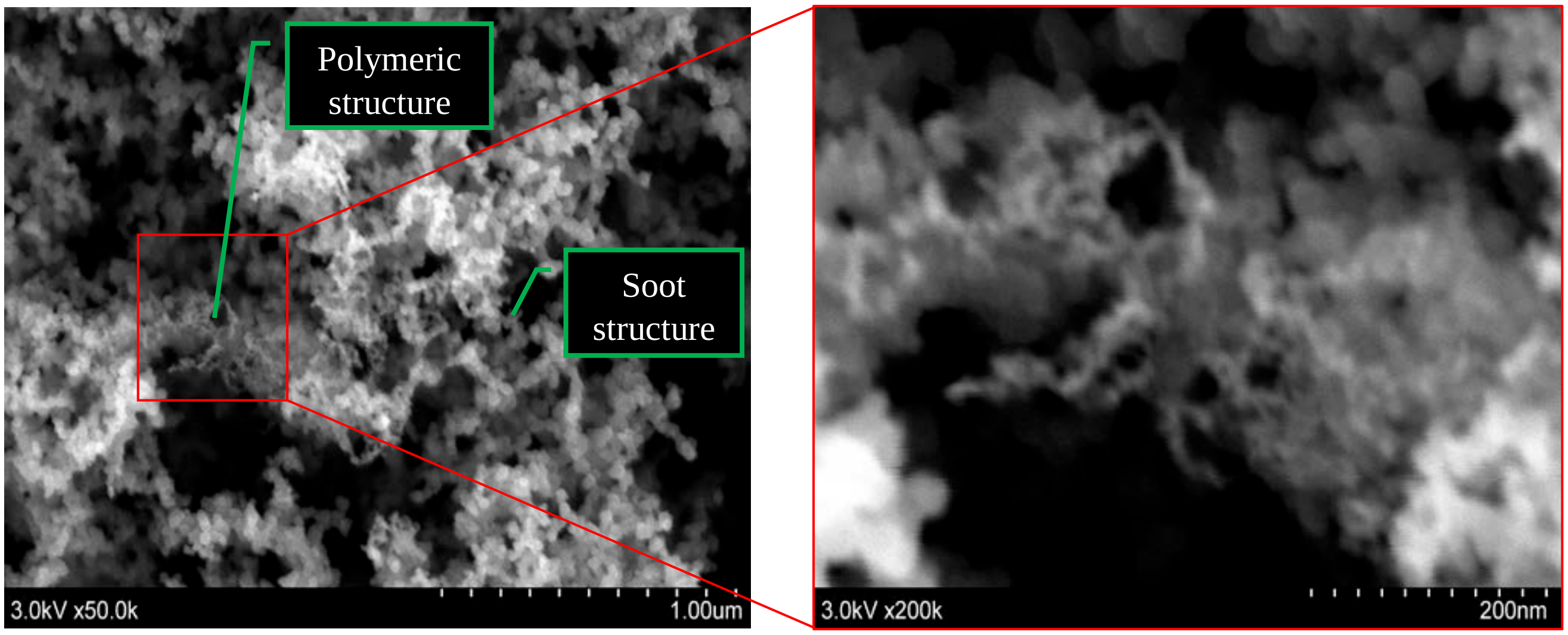

**Figure 23.** SEM images of the soot deposit if 1% PBD5k-Bakken blend soot residue, showing globular and stringy polymeric residue structure (shown in magnified images on the right) in the soot structure left behind

## 4. Conclusions:

This work explored the experimental investigation of isolated single droplet burning behavior of two US crude oils sourced from Pennsylvania and North Dakota (Bakken), and the modification of this burning behavior using polybutadiene (PBD) polymer of chain length 5,000 (PBD5k) and 200,000 (PBD200k). Treating the crude oil as a multicomponent liquid fuel, PBD was added at different mass concentrations to make fuel blends. Sub-millimeter droplets of the fuels and fuel blends were burned to completion, and the process was recorded using a CCD high speed camera and a CMOS high speed camera. CCD camera photography with a single LED as a backlight revealed time evolution of droplet area, whereas low-light photography with the same camera revealed the flame stand-off ratio. CMOS camera photography revealed ignition delay and total combustion time. Scanning electron microscopy imaging of the soot deposits left behind by the burned droplets revealed the soot structure and soot particle size. Significant changes in all these combustion and soot properties were observed. These are summarized below:

- A decrease in combustion rate is seen for Pennsylvania crude when PBD5k is added, with the largest decrease of 26% occurring at 3%w/w PBD5k. PBD200k generally increases the combustion rate in Pennsylvania crude, with a maximum increase of 27% noted at 3%w/w. A general decrease in combustion rate is noted for Bakken crude when PBD5k is added, with the largest decrease of 12% occurring at 2% w/w. PBD200k is noted to generally increase combustion rate in Bakken crude, with the largest increase of 140% noted at 3%w/w. Both PBD5k and PBD200k can be expected to decrease vapor pressure and inhibit diffusion of lighter components to the surface to decrease combustion rate, but faster thermal degradation and increased thermal conductivity of PBD200k result in an overall increase in combustion rate, thereby making PBD5k as a much more likely additive for crude oil transportation safety.
- A decrease in ignition delay is noted for Pennsylvania crude when PBD5k is added, with the largest decrease of 26% occurring at 1% w/w. Generally, addition of PBD200k to Pennsylvania crude causes an increase in ignition delay, with the largest increases of 6% occurring at 1% and 3% w/w. This can be explained by a larger ignition delay associated with PBD200k because of its larger chain length. A large increase in ignition delay for Bakken crude is noted on addition of both PBD5k and PBD200k. A maximum of 52% increase in ignition delay is observed for Bakken crude at addition of 1% PBD5k. A maximum of 42% increase can be observed in ignition delay for Bakken crude at addition of 4% PBD200k. This can be explained by Bakken crude having a lower ignition delay compared to Pennsylvania crude, and the addition of both PBD5k and PBD200k resulting in an overall decrease in ignition delay because of decreased vapor pressure.

- PBD5k causes a general increase in Pennsylvania crude total combustion time, with the greatest increase of 3.4% seen at 3% PBD5k. This is caused by the decrease in combustion rate, as noted above. Addition of PBD200k generally causes total combustion to decline with the greatest average total combustion time decrease of 15% seen at 4% PBD200k. This is caused by high intensity microexplosions at high PBD200k concentrations, which cause loss of liquid fuel. PBD5k causes a general decrease in Bakken crude total combustion time, with the greatest decrease of 8% seen at 3% PBD5k. Adding PBD200k to Bakken crude greatly decreases total combustion time with the largest average total combustion time decrease of 28% noted at 4% PBD200k. This can be explained by increased microexplosion intensity in Bakken crude when both PBD5k and PBD200k are added.
- Generally, a decrease in flame stand-off ratio (FSR) is noted when PBD5k is added to Pennsylvania crude. This is explained by the decrease in vapor pressure of crude oil with the addition of polymer, as noted above.
- SEM imaging of the soot deposits revealed that PBD5k-Bakken blends leave behind a more densely packed structure compared to pure Bakken. Individual average soot particle for PBD5k-Bakken blends was found to decrease to ~40µm from ~70µm in case of pure Bakken crude. Additionally, imaging of the soot structure surface revealed globular polymeric structure embedded in the soot residue. Ultimately, soot structure of combustion residue is determined by the chemical make-up of the fuel.

Properties such as burning rate are envisioned to be used for validation of computational and numerical modeling of multicomponent fuel droplet combustion process. Practically, decreased combustion rates achieved for PBD5k blends are expected to be useful for better crude transportation safety, whereas increased combustion rates achieved for PBD200k blends are expected to be useful for achieving better in-situ burning efficiency for crude oil spills. It is expected that present research will also aid in exploration of other polymeric additives to improve crude oil combustion safety.

## Acknowledgements:

This research is funded, in part, by the Mid-America Transportation Center via a grant from the U.S. Department of Transportation's University Transportation Centers Program, and this support is gratefully acknowledged. The USDOT UTC grant number for MATC is: 69A3551747107. The authors would also like to acknowledge use of the University of Iowa Central Microscopy Research Facility, a core resource supported by the Vice President for Research & Economic Development, the Holden Comprehensive Cancer Center and the Carver College of Medicine. We would especially like to thank Dr Jianqiang Shao for his help and patience. We would also like to thank Prof. Lynn M. Teesch and Mr. Vic R. Parcell for their help and support with the GC-MS data. The contents reflect the views of the authors, who are responsible for the facts and the accuracy of the information presented herein and are not necessarily representative of the sponsoring agencies, corporations or persons.

## Appendix A. Gas Chromatography – Mass Spectrometry (GC-MS) data for Pennsylvania and Bakken crude oils

GC-MS data for Pennsylvania (**Figure A1**) and Bakken crude (**Figure A2**) oils was generated at the High Resolution Mass Spectrometry Facility (HRMSF) located at the University of Iowa Department of Chemistry. Bakken crude oil GC-MS data has been previously reported by Singh *et al.* [15]. The column used was a 30 m DB-5MS, 0.25mm diameter and 0.25µm film thickness. The temperature ramp started at 50 ºC and held for 1 min. It was then increased at 10 ºC/minute until 320 ºC and then held for 5 minutes. The numbers on top of the peaks in the chromatogram are retention time, area, and response height. The results reveal that both crudes are rich in low- and medium-boiling components, with Bakken being richer in low-boiling components compared to Pennsylvania crude.

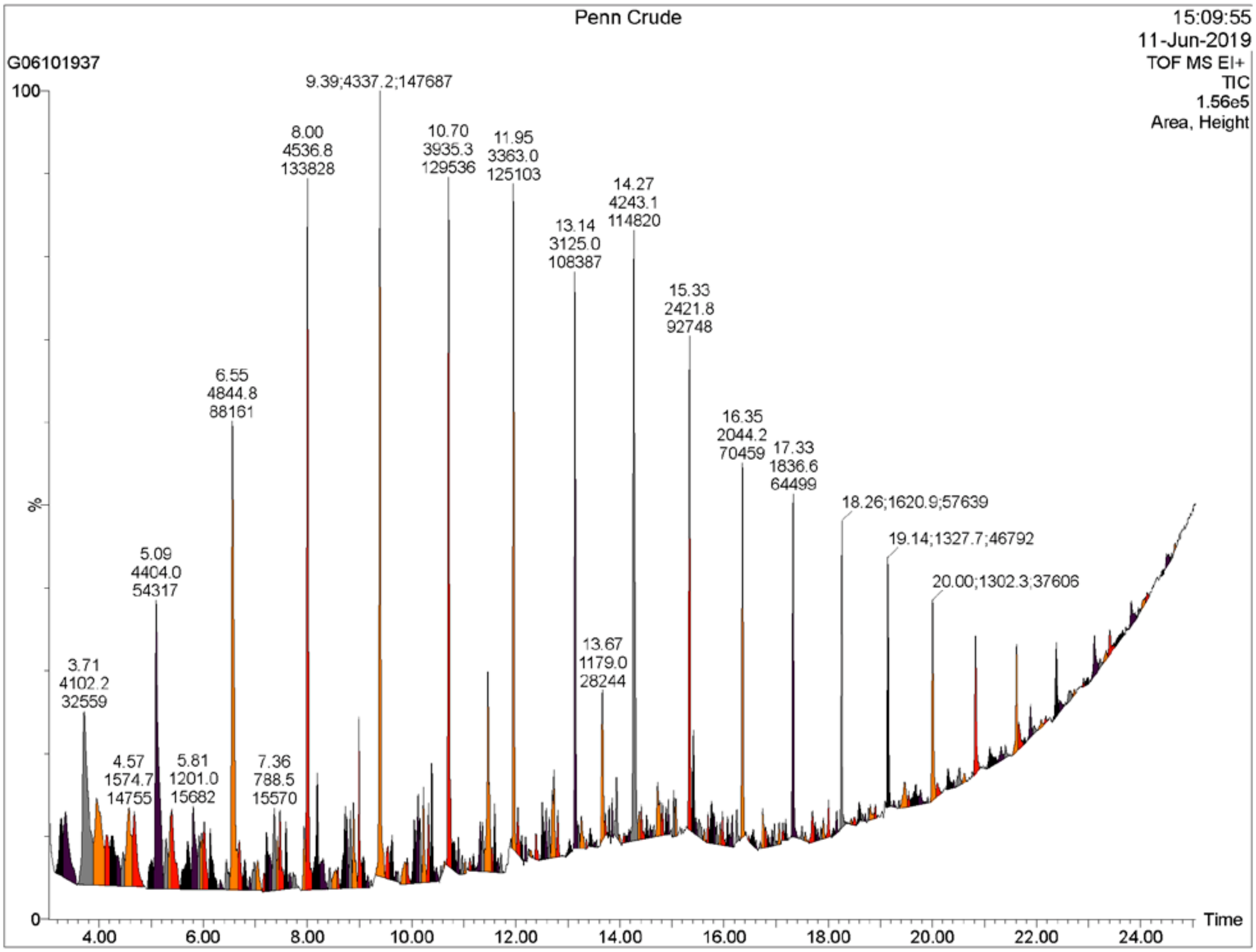

**Figure A1.** Gas Chromatography – Mass Spectrometry (GC-MS) data for Pennsylvania crude oil.

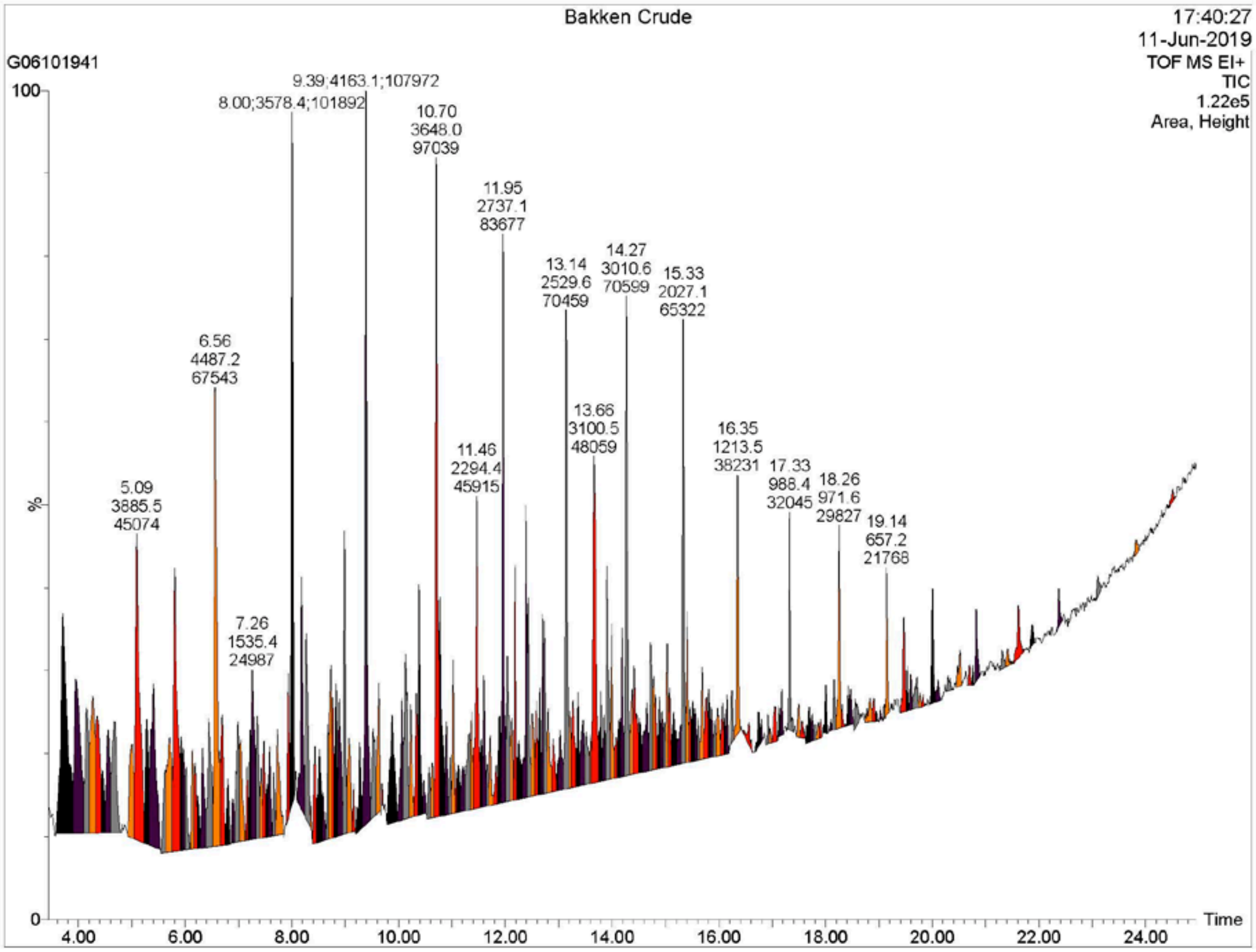

**Figure A2.** Gas Chromatography – Mass Spectrometry (GC-MS) data for Bakken crude oil.